\documentclass{emulateapj}
\usepackage{apjfonts}

\newcommand{\figexpand}{\epsscale{1.15}}

\newcommand{\etal}{et al.}

\newcommand{\mtrans}{M_{tr}}
\newcommand{\mhaloquench}{M_{\rm crit}}

\shorttitle{Co-evolution of Mergers, Quasars, and Red Galaxies}
\shortauthors{Hopkins \etal}
\slugcomment{ApJ 2006}
\begin{document}

\title{Observational Evidence for the Co-evolution of Galaxy Mergers,
Quasars, and the Blue/Red Galaxy Transition}
\author{Philip F. Hopkins\altaffilmark{1}, 
Kevin Bundy\altaffilmark{2}, 
Lars Hernquist\altaffilmark{1}, 
\& Richard S. Ellis\altaffilmark{2}
}
\altaffiltext{1}{Harvard-Smithsonian Center for Astrophysics, 
60 Garden Street, Cambridge, MA 02138}
\altaffiltext{2}{105-24 Caltech, 1201 E.\ California Blvd., 
Pasadena, CA 91125}

\begin{abstract}
We compile a number of recent observations to estimate
the time-averaged rate of formation or ``buildup'' of red sequence galaxies, as 
a function of mass and redshift. Comparing 
this with the mass functions of mergers and quasar hosts, and independently comparing 
their clustering properties as a function of redshift, 
we find that these populations trace the same mass distribution, with the 
same characteristic masses and evolution, in the redshift interval
$0<z\lesssim1.5$. Knowing one of the 
quasar, merger, or elliptical mass/luminosity functions, it is possible to predict the others. 

Allowing for greater model dependence, we 
compare the rate of early-type ``buildup'' with the implied 
merger and quasar ``triggering'' rates as a function of 
mass and redshift, and find agreement. 
We show that over this redshift range, 
observed gas-rich merger fractions can account for the entire bright 
quasar luminosity function, and buildup of the red sequence at all but 
the highest masses at low redshift ($\gtrsim10^{11}\,M_{\sun}$ at $z\lesssim0.3$) 
where dissipationless ``dry'' mergers appear to dominate. 
This supports a necessary prediction of theories which postulate that
mergers between gas-rich blue galaxies produce ellipticals with an
associated phase of bright quasar activity, after which the remnant becomes red.
All of these populations, regardless of sample selection, trace a similar characteristic 
``transition'' mass reflecting the 
characteristic mass above which the elliptical population is mostly 
($\gtrsim50\%$) assembled at a given redshift, which increases with redshift over the observed range 
in a manner consistent with previous suggestions that ``cosmic 
downsizing'' may apply to red galaxy {\em assembly} as well as star formation. 
We show that these mass distributions as a function of redshift do not 
uniformly trace the all/red/blue galaxy population, ruling out models in 
which quasar activity is generically associated with either star formation or 
is long-lived in ``old'' systems. 
\end{abstract}


\section{Introduction}
\label{sec:intro}

Observations motivate the notion of ``cosmic downsizing'' 
\citep[as coined by][]{Cowie96},
with the global star formation rate declining rapidly below $z\sim2$,
and the sites of galactic star formation shifting to smaller masses
at lower redshift. Moreover, galaxy surveys such as SDSS, COMBO-17,
and DEEP2 demonstrate that the color distribution of galaxies is
bimodal \citep[e.g.,][]{Strateva01,Balogh04}, and that this bimodality
extends at least to $z\sim1$ \citep[e.g.,][]{Bell04,Faber05}.

It is 
increasingly established that high
mass, red elliptical galaxies have older 
stellar populations than smaller spheroids \citep[e.g.,][]{Caldwell03,Nelan05,Gallazzi06}.
But, many studies 
also see a significant population of massive/luminous galaxies 
in place (i.e.\ assembled) by $z\sim2$ \citep[e.g.,][and references therein]{Papovich06,Renzini06}, 
with measurements 
of galaxy stellar mass functions (MFs) and luminosity functions at 
redshifts $0<z<2$ favoring either a uniform increase or buildup 
in the numbers of early-type (``red sequence,'' RS) 
galaxies \citep[e.g.,][]{Bell04,Faber05} or an anti-hierarchical scenario 
in which this ``buildup'' at $z\lesssim1$ occurs primarily at the
low-mass end of the RS \citep{Bundy05.full,Bundy05.masslimit,Zucca05,Yamada05, 
Borch06,Franceschini06,Pannella06,Cimatti06,Brown06}. The blue, disk dominated, 
star forming galaxy mass function (dominant at low mass), meanwhile, remains 
relatively constant, or perhaps declines to $z=0$. 
As a consequence, the ``transition mass,'' above which the red
galaxy population dominates the galaxy MF, decreases with time, 
tracing this downsizing trend. There is evidence for some evolution at the highest 
masses as ellipticals grow by spheroid-spheroid or ``dry'' mergers \citep{vanDokkum05,Bell06}, 
but this, by definition, proceeds strictly hierarchically, and 
{\em cannot} account for the movement of mass onto the RS in the 
first place or any buildup in the number density of low-mass ellipticals. 

Meanwhile, the discovery of
tight correlations between the masses of central supermassive black
holes (BHs) in galaxies and the bulge or spheroid stellar mass
\citep{Magorrian98}, velocity dispersion \citep{Gebhardt00,FM00} 
or concentration \citep{Graham01} 
implies that the formation of galaxies and BHs
must be linked.  Moreover, the
evolution of the quasar luminosity function (QLF) shows a 
sharp decline after $z\sim2$, with the density of
lower-luminosity AGN peaking at low redshift \citep[e.g.,][and
references therein]{HMS05}. To the extent that BH assembly traces 
galaxy assembly (i.e.\ there is weak evolution in the BH-host mass 
relation, as observed to at least $z\gtrsim1$ by e.g.\ \citet{Shields03,AS05a,Peng06}), 
this implies early {\em assembly} times ($z\gtrsim1$) for many of 
the most massive systems containing $M_{\rm BH}\gtrsim10^{8}\,M_{\sun}$ BHs. 

A number of theoretical models have been proposed to explain the 
evolution of these populations with redshift, and their 
correlations with one another 
\citep[e.g.,][]{KH00,Somerville01,WyitheLoeb03,Granato04,Scannapieco05,
Baugh05,Monaco05,Croton05,H06b,H06c,H06d,Cattaneo06}. In many of these 
models, the merger hypothesis \citep{Toomre77} provides 
a potential physical mechanism linking galaxy star formation, morphology, and 
black hole evolution and explaining these various manifestations of 
cosmic ``downsizing.'' In this scenario, gas-rich galaxy mergers channel 
large amounts of gas to galaxy centers
\citep[e.g.,][]{BH91,BH96}, fueling powerful starbursts 
\citep[e.g.,][]{MH94,MH96}
and buried BH growth \citep[e.g.,][]{Sanders88,BH92} until the BH
grows large enough that feedback from accretion rapidly unbinds and heats the
surrounding gas \citep{SR98}, leaving an elliptical galaxy
satisfying observed correlations between BH and spheroid mass. 
Major mergers rapidly and efficiently exhaust the cold gas reservoirs of the 
progenitor systems, allowing the remnant to rapidly redden with a low 
specific star formation rate, with the process potentially accelerated by the 
expulsion of remnant gas by the quasar \citep[e.g.,][]{SDH05a}. This naturally 
explains the observed close association between the elliptical 
and red galaxy populations \citep[e.g.,][]{Kauffmann03}.

In a qualitative sense, the evolution of the characteristic mass at which these 
processes occur can be understood as follows.
Mergers proceed efficiently at high redshift, occurring most rapidly in the regions of 
highest overdensity corresponding to the most massive galaxies, building up the
high-mass elliptical MF.  However, once formed these galaxies are ``dead'',
and mergers involving gas-rich galaxies must transition to lower masses. 

Recent hydrodynamical simulations, incorporating star formation, supernova
feedback, and BH growth and feedback \citep{SDH05b} make it possible
to study these processes self-consistently and have lent support to this 
general picture.  Mergers with BH feedback
yield remnants resembling observed ellipticals in their correlations
with BH properties \citep{DSH05}, scaling relations
\citep{Robertson05b}, colors \citep{SDH05a}, and morphological and
kinematic
properties \citep{Cox06a,Cox06b}.  The quasar activity excited through such
mergers can account for the QLF and a wide range of quasar properties
at a number of frequencies \citep{H05a,H06b}, and with such a detailed
model to ``map'' between merger, quasar, and remnant galaxy
populations it is possible to show that the buildup and statistics of
the quasar and red galaxy populations are consistent and can be used
to predict one another \citep{H06c}.

However, it is by no means clear whether this is, in fact, the dominant 
mechanism in the buildup of early-type populations and quasars and their 
evolution with redshift. 
For example, many semi-analytic models 
incorporate quasar triggering/feedback and morphological transformation by mergers 
\citep{KH00,Volonteri03,Volonteri06,
WyitheLoeb03,Somerville04a,Monaco05,Bower06,Lapi06,Menci06}.
However, some models tie quasar activity directly 
to star formation \citep[e.g.,][]{Granato04}, implying it will evolve in a manner 
tracing star-forming galaxies, with this evolution and the corresponding downsizing 
effect roughly independent of mergers and morphological 
galaxy segregation at redshifts $z\lesssim2$. 
Others invoke post-starburst AGN feedback to suppress 
star formation on long timescales and at relatively low 
accretion rates through e.g.\ ``radio-mode'' feedback 
\citep{Croton05}, which, if this is also associated with optical QSO modes, 
would imply quasars should trace the established ``old'' red galaxy 
population at each redshift. 
There are, of course, other sources of feedback, with galactic superwinds 
from star formation presenting an alternative means to suppress subsequent star 
formation, although the required wind energetics are sufficiently high 
to prefer a quasar-driven origin \citep[e.g.,][]{Benson03}. Several models 
invoke a distinction between ``hot'' and ``cold'' accretion modes 
\citep{Birnboim03,Keres05,Dekel06}, in which new gas cannot cool into a 
galactic disk above a critical dark matter halo mass, potentially supplemented by 
AGN feedback \citep{Binney04}, as the dominant distinction between 
the blue cloud and red sequence, essentially independent of effects on 
scales within galaxies.

It is also important to distinguish the processes which may be associated 
with the initial movement of galaxies onto the red sequence from their subsequent 
evolution. Once morphologically transformed by a gas-rich merger, for example, 
mass can be moved ``up'' the RS (galaxies increased in mass)
by gas-poor mergers, but it cannot be {\em added} to the red 
sequence in this manner. It also remains an important cosmological question to 
understand how, once formed, 
further growth of ellipticals by accretion or ``cooling flows'' may be halted. The models 
above invoke various feedback processes, including 
``radio mode'' activity \citep[e.g.,][]{Croton05}, 
cyclic quasar or starburst-driven feedback \citep{Somerville01,Granato04,Binney04,Monaco05}, 
massive entropy injection from a single quasar epoch \citep[e.g.,][]{WyitheLoeb03,Scannapieco04}, 
and ``hot mode'' accretion \citep{Birnboim03} 
to address this problem. Although 
critical to our understanding of galaxy formation, these processes must operate 
over timescales of order the Hubble time for all massive galaxies once formed, and therefore 
are {\em not} necessarily associated with the {\em addition} of mass to the red sequence. 
As such, the details of these long-term suppression mechanisms  
should be studied in different (e.g.\ already formed elliptical) populations, and are 
outside the scope of this paper. 

Observationally, it is still unclear whether mergers can account for 
the the buildup of elliptical and/or quasar 
populations \citep[see, e.g.,][and references therein]{Floyd04,RJ06,Lotz06b}. 
Even within the context of the merger hypothesis, the
relative importance of dissipational (gas rich, disk) vs.\ dissipationless (gas poor, 
spheroid-spheroid) mergers is unclear \citep[e.g.,][]{vanDokkum05,Bell06}, 
although all measurements agree that the ``dry'' merger rate is much less 
than the gas-rich merger rate at all observed redshifts \citep{Bell06,Lotz06b,Bell06b}. This is
essentially related to the critical question of whether the buildup of the 
red sequence and elliptical populations is dominated by the formation or movement of ``new'' 
early-type galaxies onto that sequence or instead by the hierarchical assembly 
of small ``seed'' early-types and substructure formed at high redshift (which will also 
not trigger quasar activity). 

Fundamentally, it is not clear and has not yet been tested whether the observed downsizing trends 
in the transition mass, galaxy stellar populations, quasars, and other populations 
are in fact {\em quantitatively} the same trend, or merely {\em qualitatively} 
similar. This represents a key test which can distinguish between several of 
the various scenarios above. 
Attempting to predict the values of this transition mass in an a priori 
cosmological manner is inherently model dependent and, at least 
at low redshift, degenerate between the various models described above. However, 
if mergers are indeed the critical link in the process causing the flow 
of galaxies from the blue to red sequence and triggering quasar activity, then 
it is a strong prediction of these theories, and
specifically the modeling of Hopkins et al.\ (2005a-d,\,2006a-d) that the same
mergers are responsible for the bulk of the bright quasar population
and the buildup of the new mass on the 
red sequence at each redshift. In other words, these downsizing 
trends must quantitatively reflect one another. 

In this picture, the ``transition mass'' ($\mtrans$) may represent the
``smoking gun'' of mergers causing the flow of galaxies from the blue
to red sequence. Therefore, to
the extent that $\mtrans$ traces the mass at which the red sequence is
being ``built'' at some $z$, it should also trace the characteristic
mass of star-forming galaxies merging at that time, and the
characteristic mass of galaxies hosting quasars which are initially
triggered by those mergers. Of particular interest, the empirical test of this association does 
not require the adoption of some a priori model for galaxy formation. 

Here, we consider the observed $\mtrans$
over the interval $0<z<2$, and compare it to the characteristic masses of
quasar hosts and merging galaxies over the same range in redshift. We
demonstrate that they appear to be evolving in a manner consistent
with a merger-driven unification model of quasars, interacting
galaxies, and the red galaxy population. 
Note that we use the term ``quasar'' 
somewhat loosely, as a proxy for high-Eddington ratio accretion 
inevitably caused by gas-rich mergers, although there may be
other triggering mechanisms as well 
\citep{Sanders88,Alexander05a,Alexander05b,Borys05,H06b}. 
Such activity will of course be significantly weaker in small systems (especially 
those typical of local ULIRGs) and may not technically qualify as a 
``classical'' optical quasar \citep{H05b}, but this distinction 
is essentially arbitrary and has little impact on our analysis 
\citep[see also][]{H06d}. 

We adopt a $\Omega_{\rm M}=0.3$, $\Omega_{\Lambda}=0.7$,
$H_{0}=70\,{\rm km\,s^{-1}\,Mpc^{-1}}$ cosmology. All stellar masses 
are rescaled to a \citet{Salpeter55} IMF.

\section{The Transition Mass And Buildup of Early-Type Populations}
\label{sec:mtrans}

\subsection{Defining the ``Transition'' Mass}
\label{sec:defns}

Various studies have used different definitions and terms for 
the mass which separates the dominance of old, red, low-SFR elliptical 
galaxies from that of young, blue, star-forming disk galaxies. It is also 
possible to divide the galaxy population 
along any one of those quantities. Although it has been 
established in a number of observational studies that the 
galaxy population is bimodal with respect to 
color, specific star formation rate, and morphology 
\citep[e.g.,][]{Strateva01,Kauffmann03,Balogh04,Driver05}, and that this bimodality
extends at least to $z\sim1$ \citep[e.g.,][]{Bell04,Faber05}, it is 
still possible that the various definitions used to separate 
these bimodal distributions could result in a systematically 
different ``separation point.''

In what follows, we consider several definitions of the ``transition'' 
galaxy stellar mass in terms of the MFs ($\phi(M)$) of early and late type systems:
the \citet{Bundy05.full} transition mass $\mtrans$ at which the density of 
early and late type systems are equal,
\begin{equation}
\phi_{\rm early}(\mtrans,\,z) = \phi_{\rm late}(\mtrans,\,z),
\end{equation}
the  \citet{Bundy05.masslimit} ``quenching'' mass $M_{Q}$ at which the contribution of late types 
to the total mass function cuts off, 
\begin{equation}
\phi_{\rm late}(M,\,z) = \phi_{\rm all}(M,\,z)\,\exp{(-M/M_{Q})},
\end{equation}
and the \citet{Cimatti06} ``downsizing'' mass $M_{50}$ above which 
$50\%$ of the $z=0$ RS MF has been assembled by a given 
redshift, 
\begin{equation}
%\int^{\infty}_{M_{50}}\phi_{\rm early}(M,\,z) = 0.5\,\int^{\infty}_{M_{50}} \phi_{\rm early}(M,\,0)
\phi_{\rm early}(M>M_{50},\,z) \ge 0.5\,\phi_{\rm early}(M>M_{50},\,z=0).
\end{equation}
For each, we consider a division between early and late types 
defined by either a color (i.e.\ 
separating galaxies on the redshift-dependent RS from the ``blue cloud''; see e.g.\ \citet{Bell04}), 
star formation rate (generally from detailed population synthesis modeling, although we 
include looser emission/absorption galaxy spectral type separations), 
or morphology (either by visual morphological identification or 
concentration/surface brightness) criteria. 
Above these masses, early type galaxies 
in the red, low-SFR, high-concentration, morphologically elliptical 
``half'' of the bimodal distribution dominate the total galaxy MF, 
and below, late type galaxies in the 
blue, high-SFR, low-concentration, morphologically disk-like or irregular 
bimodal ``half'' dominate (generally true also for $M_{50}$, though it is technically 
independent of $\phi_{\rm late}$).

\begin{figure*}
    \centering
    \figexpand
    \plotone{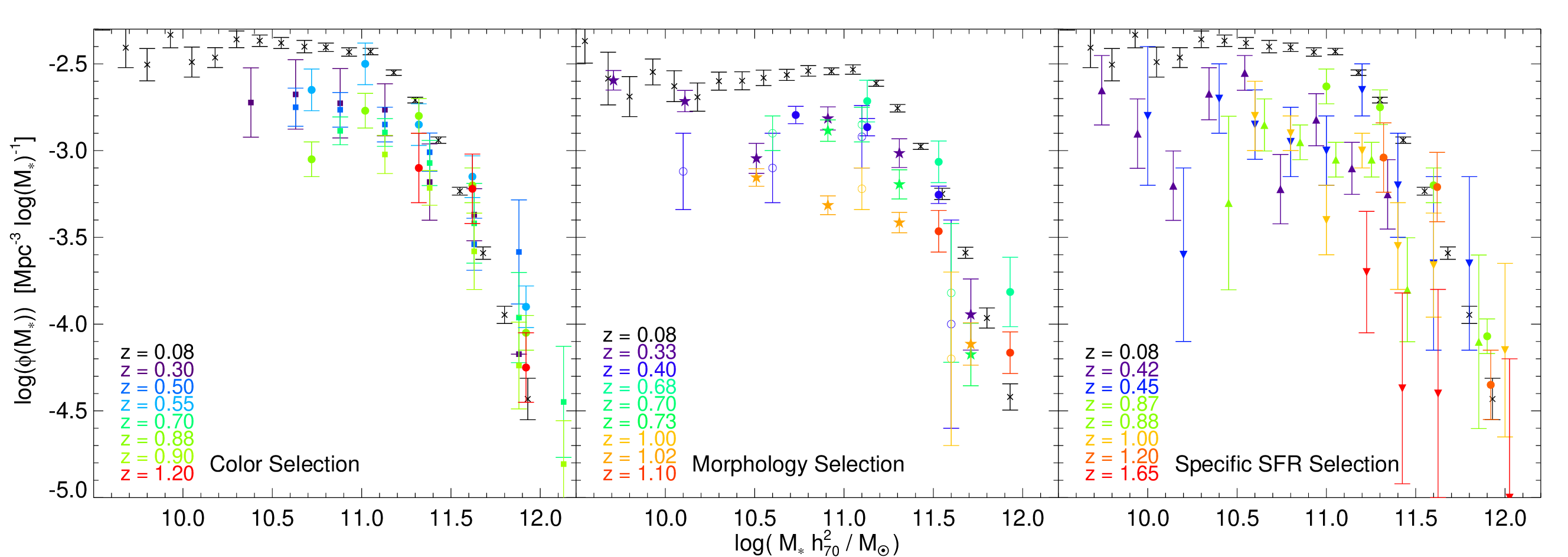}
    \caption{Mass functions of early type galaxies selected by 
    color, morphology, or specific star formation rate, from 
    \citet[][$\times$'s]{Bell03}, 
    \citet[][squares]{Borch06}, 
    \citet[][filled circles]{Bundy05.full,Bundy05.masslimit}, 
    \citet[][open circles]{Pannella06},
    \citet[][stars]{Franceschini06}, 
    \citet[][inverted triangles]{Fontana04}, and 
    \citet[][triangles]{Pozzetti03}. 
    Points are colored by redshift, as labeled, 
    and shown only above the quoted completeness limits of each study. 
    The data have been converted to our adopted cosmology 
    and masses rescaled to a \citet{Salpeter55} IMF. 
    Comparison suggests that ``cosmic downsizing'' may apply in some 
    sense to early-type galaxy assembly, as well as star formation histories. 
    \label{fig:oplot.mfs}}
    %\epsscale{1.0}
\end{figure*}

We determine these masses from a number of compiled type-separated 
mass functions, shown in Figure~\ref{fig:oplot.mfs} (for clarity, just the 
early-type MFs are shown). Data in all cases 
are converted to our adopted cosmology and rescaled to a \citet{Salpeter55} IMF. 
At all redshifts, only points above the quoted completeness limits of each 
study are shown. Errors are as published, and generally account for 
cosmic variance similar to e.g.\ \citet{Somerville04b}. 
At $z=0$, we generally adopt the local MF determinations from 
\citet{Bell03} from 2MASS+SDSS observations, but find no change 
in our results considering e.g.\ the 2MASS+2dFGRS \citet{Cole01} determination. 
At higher redshifts, our compilation includes 
\citet[][K20; $z<1.3$]{Pozzetti03}, \citet[][K20; $z<2.0$]{Fontana04}, 
\citet[][GOODS; $z<1.4$]{Bundy05.full} 
and \citet[][DEEP2; $z<1.4$]{Bundy05.masslimit}, 
\citet[][COMBO-17; $z\leq1.0$]{Borch06}, 
\citet[][CDF-S; $z<1.4$]{Franceschini06}, and 
\citet[][FORS Deep+GOODS-S; $z\leq1.15$]{Pannella06}. Each of these 
considers the separate MFs of early and late types, divided according 
to at least one of the criteria above. Although many of 
these surveys cover a small area and thus may be subject to 
significant cosmic variance, our conclusions are unchanged if we restrict 
ourselves to the 
largest fields, and are in fact most robust for the 
wide-field surveys from COMBO-17 ($0.78\,{\rm deg}^{2}$) and 
DEEP2 ($3.5\,{\rm deg}^{2}$). 

%\clearpage
%\clearpage

\begin{figure*}
    \centering
    \figexpand
    \plotone{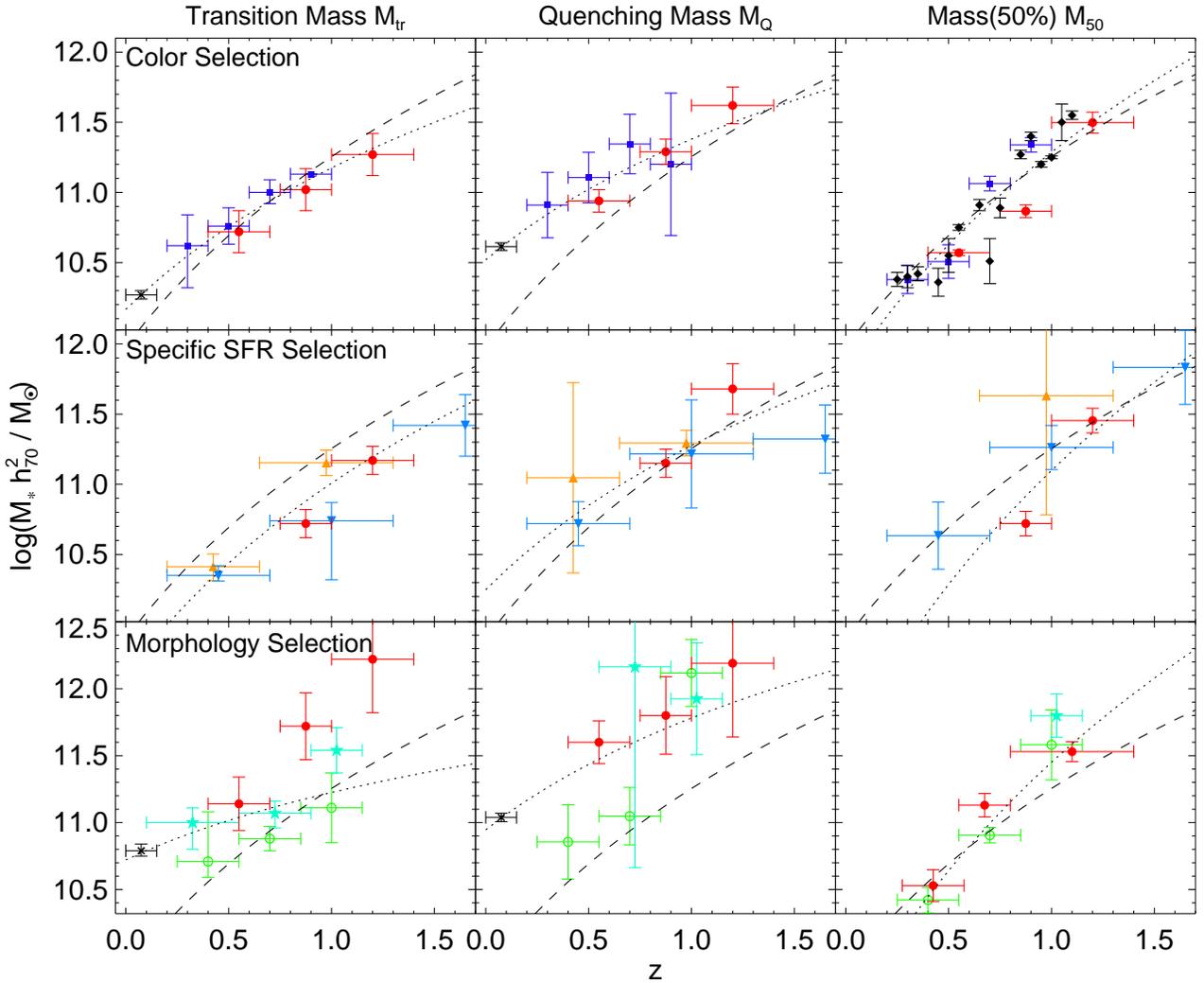}
    \caption{Evolution with redshift of the ``transition'' or ``downsizing'' mass, 
    at which objects may be in transition from the blue cloud to the red sequence 
    (i.e.\ the characteristic mass at which elliptical populations may be ``building up'') 
    with different sample selections and definitions of this mass. 
    {\em Left:} $\mtrans$, 
    the mass above which ellipticals dominate the cumulative galaxy mass function, 
    with galaxy types separated by color selection (upper), 
    specific star formation rate or spectral (absorption/emission) fitting (middle), 
    and morphological selection (lower). 
    {\em Center:} $M_{Q}$, the mass at which the contribution of late-type galaxies 
    cuts off, with the same sample definitions as for $\mtrans$. 
    {\em Right:} $M_{50}$, the mass above which the early-type mass function 
    at $z$ is $\ge50\%$ assembled relative to the \citet{Bell03} mass function 
    at $z=0$ (i.e.\ $\phi(M>M_{50},\,z) \ge 0.5\,\phi(M>M_{50},\,z=0)$). 
    Data are shown from the mass functions in Figure~\ref{fig:oplot.mfs}, 
    in the same point style, color-coded by the observed sample: 
    \citet[][black $\times$'s]{Bell03}, 
    \citet[][purple squares]{Borch06}, 
    \citet[][red filled circles]{Bundy05.full,Bundy05.masslimit}, 
    \citet[][green open circles]{Pannella06},
    \citet[][cyan stars]{Franceschini06}, 
    \citet[][blue inverted triangles]{Fontana04}, and 
    \citet[][orange triangles]{Pozzetti03}. 
    We also consider $M_{50}$ calculated in \citet[][black diamonds]{Cimatti06} from the 
    luminosity functions of 
    \citet{Bell04,Willmer05,Faber05,Yamada05}, 
    using the redshift-dependent mass-to-light ratios estimated from 
    fundamental plane studies. 
    Dotted lines in each panel show the best-fit trend of the form 
    $M_{\ast} \propto (1+z)^{\kappa}$. Dashed lines (identical in all panels) 
    show a cumulative best-fit 
    to the $M_{50}$ data from all samples. 
    Although there are systematic 
    factor $\sim2$ normalization offsets between different methods which caution against 
    mixing definitions, the various methods all trace a similar mass.
    Regardless of the characteristic mass definition, 
    the sample survey, or 
    the method of type segregation of the samples, a similar trend with 
    redshift is recovered in each case. 
    \label{fig:compare.methods}}
    %\epsscale{1.0}
\end{figure*}

Figure~\ref{fig:compare.methods} shows resulting ``transition'' mass from 
these studies as a function of redshift, from the different definitions and galaxy type segregation 
methods above. These samples generally derive masses from 
optical+near-IR spectral and photometric fitting. To compare, we also 
consider the evolution in $M_{50}$ (color-selected) determined 
by \citet{Cimatti06}. They compile the $B$-band early-type (red) 
galaxy luminosity functions from 
COMBO-17 \citep{Bell04}, DEEP2 \citep{Willmer05,Faber05}, and the 
Subaru/{\em XMM-Newton} Deep Survey \citep{Yamada05} at $z=0-1.15$ 
(typical $\Delta z\approx0.05$ between luminosity functions), and use 
the redshift-dependent evolution of $B$-band mass-to-light 
ratios determined from fundamental plane studies \citep[e.g.,][roughly 
similar to mean formation redshifts $z_{f}\sim3-4$]{vanDokkumStanford03,vdW05,
Treu05,diSeregoAlighieri05,Renzini06} to correct these to a $z=0$ equivalent 
$B$ luminosity and mass (given the $z=0$ mass-dependent 
$B$/$g$-band $M/L$ ratios from \citet{Bell03}). 
We fit a simple relation of the form 
\begin{equation}
%\log(M)=\log(M_{0})+k\,z
M_{\ast} \propto (1+z)^{\kappa}
\end{equation}
in each panel (for illustrative purposes only, we do not intend for this 
to be considered a rigorous estimate of the evolution in these characteristic 
masses).

%\clearpage
%\clearpage

Despite the different selection and type separation 
methods and definitions of a characteristic 
mass, a nearly identical trend with redshift is recovered in every case. In fact, 
the best-fit slopes $\kappa$ for most selection methods and definitions are statistically 
indistinguishable from the cumulative best-fit slope, and 
several of the definitions agree nearly exactly in Figure~\ref{fig:compare.methods}. 
It is also reassuring that the indirect estimates from optical luminosity functions, 
which generally involve the largest samples and most finely probe 
the redshift evolution of $M_{50}$, agree well at all redshifts with the MF estimates.
This suggests that the trend with redshift is real, and that it is independent of 
the potential systematics in sample selection, as these systematics can be quite different 
for the various criteria shown. 

The absolute normalization of the ``transition'' 
mass does depend systematically on the definition chosen. For example, 
same separation by color or SFR gives a systematic factor $\sim2$ lower 
mass than separation by morphology, and $M_{Q}$ is systematically 
higher than $M_{tr}$ by about the same factor for all separation methods 
\citep[see also][]{Bundy05.masslimit}. The 
systematic difference in $M_{tr}$ and $M_{Q}$ can be understood
as a consequence of their definitions (essentially one could define 
arbitrary Schechter functions for early and late-type systems, and as long 
as the early-type function has a shallower slope and larger $M_{\ast}$, 
this small systematic offset in the two measurements would be guaranteed). 
The systematic offset between color/SFR and morphological selection is best 
demonstrated in the detailed comparison of local color and morphologically 
selected MFs in e.g.\ \citet{Bell03}.  These authors
find that although the different selection methods 
preserve the same qualitative behavior, and result in early and late-type 
samples which are identical in $\sim80-90\%$ of the included galaxies, the color 
criterion does result in a slightly larger number of early-type systems (probably owing to 
the large scatter in blue galaxy colors, with a non-negligible highly dust-reddened 
population), which will push the characteristic separation mass slightly lower. 
This may also explain why there 
appears to be a larger scatter between samples at a given redshift in the 
morphologically-defined $M_{tr}$ and $M_{Q}$, as such an effect will 
be sensitive to a given sample's resolution and imaging depth.
There may also be an interesting timescale effect, as discussed in 
\citet{Bundy05.masslimit}, if galaxies redden onto the red sequence somewhat more rapidly 
than they morphologically relax following mergers (perhaps 
suggesting different mechanisms for morphological and color transformation). However, that 
$M_{50}(z)$ (depending only on $\phi_{\rm early}$) 
is similar regardless of selection method suggests that these 
differences may be an artifact of the selection/identification of 
{\em blue} galaxies. In any case, further detailed 
study of these intermediate objects and comparison between different samples 
is needed to understand these differences.

These systematic distinctions caution against mixing definitions in 
determining the redshift evolution of these masses. However, 
for our purposes, the systematic normalization scatter of a factor $\sim2$ is 
not large -- this is comparable to 
the inherent ambiguity in defining a ``characteristic mass'' of any population 
(e.g.\ Schechter $M_{\ast}$). As long as we are careful about the relatively small normalization 
offsets between selection criteria, we can safely compare 
the ``transition'' mass and its evolution with redshift to the masses of other populations.

\subsection{The Buildup of Ellipticals and the Physical 
Significance of the Transition Mass}
\label{sec:significance}

It has been suggested \citep[e.g.][]{Bundy05.full} that the 
``transition'' mass may represent the mass at which the 
early-type MF is ``building up'' at each redshift, in the sense 
that ``new'' spheroids are being added to the RS MF at this 
mass. If cosmic downsizing applies to 
galaxy assembly to any extent -- i.e.\ this ``building up'' extends 
at higher masses at higher redshift (or ``peaks'' at 
lower redshift for lower masses), then this implies a shift of $M_{50}$ 
to lower mass at lower redshift \citep{Cimatti06}. To the extent that 
the shape of the late-type MF does not dramatically change over this 
redshift interval \citep[observed in the samples above and e.g.][]{Feulner03,Drory04}
this will, by definition, manifest in a similar evolution of $M_{tr}$ and $M_{Q}$. 
In this scenario, then, $M_{50}$ is directly tied to ``downsizing'' in the 
early-type MF.

\begin{figure*}
    \centering
    %\epsscale{1.1}
    \plotone{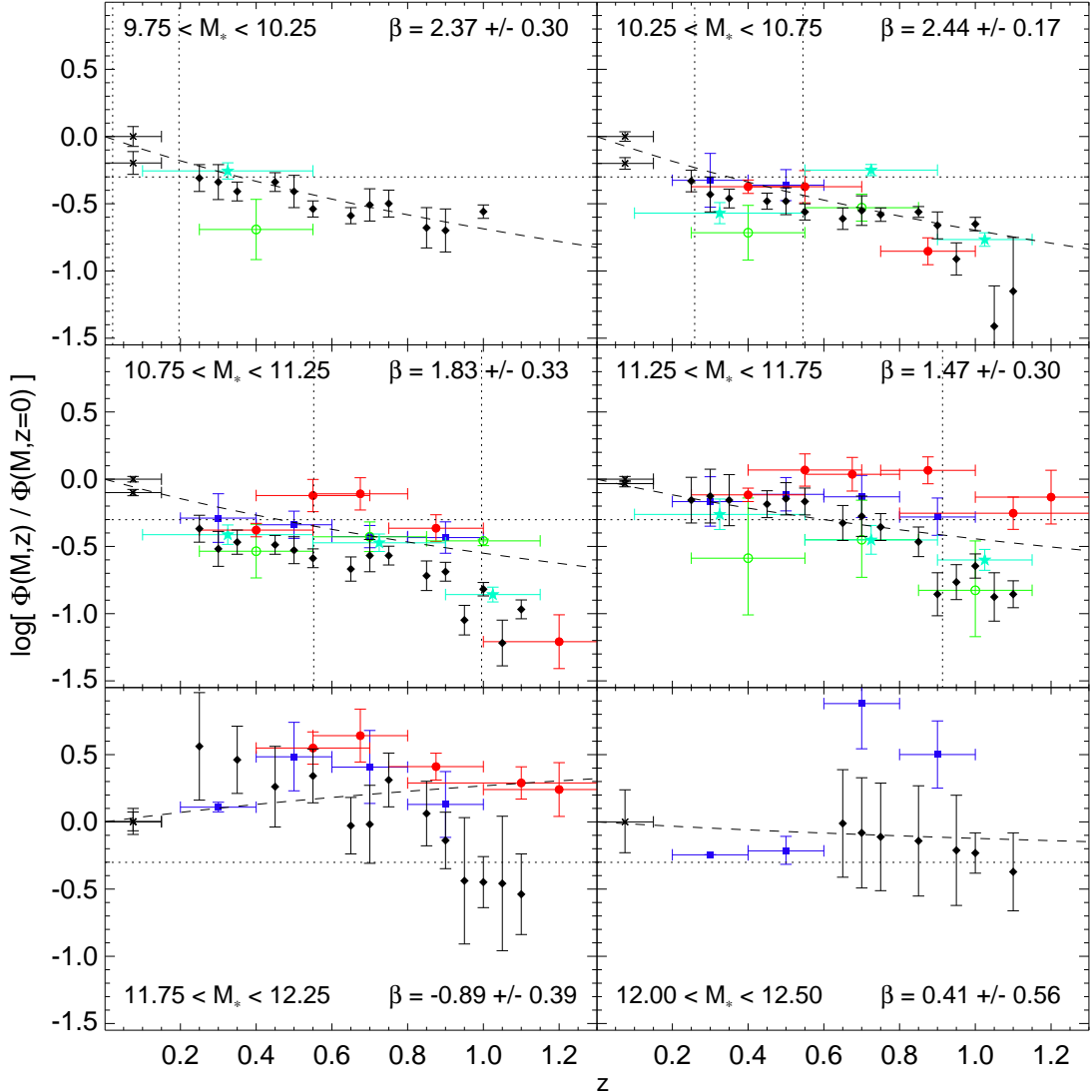}
    \caption{Number density of early-type galaxies in different mass bins as a function of 
    redshift, relative to that at $z=0$ from \citet[][color-selected]{Bell03}, 
    from the samples in Figure~\ref{fig:compare.methods} (same style). Dotted horizontal 
    line shows $50\%$ of the $z=0$ value, vertical horizontal lines show the redshifts 
    at which the lower and upper limits of each mass bin correspond to the best-fit 
    $M_{50}(z)$ (dashed line in Figure~\ref{fig:compare.methods}). Dashed lines 
    in each panel show the best-fit trend of the form $\propto (1+z)^{-\beta}$, with the 
    labeled $\beta$. Although there is considerable variance at high masses, 
    a mass-independent galaxy density evolution $\beta$ can be ruled out at 
    $\sim7\,\sigma$. A steeper $\beta$ at low mass implies that a large fraction of 
    these galaxies are added at lower redshift than galaxies of higher mass.         
    \label{fig:n.m.t}}
    %\epsscale{1.0}
\end{figure*}

Figure~\ref{fig:n.m.t} considers the evolution with redshift of the number 
density of early-type galaxies of a given $M_{\ast}$, from $9.75\leq\log{(M_{\ast}/M_{\sun})}\leq12.5$. 
At low $M_{\ast}$, the number density declines steeply with $z$ (roughly 
$\propto(1+z)^{-\beta}$ with $\beta=2.37$). Although the statistics are poor and 
variance large at high $M_{\ast}$, there is a significant trend ($\sim7\sigma$) 
for a shallower decline in number density at higher masses ($\beta\sim0$, i.e.\ 
little evolution in number density, at $M_{\ast}\gtrsim10^{12}\,M_{\sun}$). 
Above $z\sim1$, the various samples plotted begin to disagree, and the 
$\propto(1+z)^{-\beta}$ functional form may not be a good approximation, 
so we reconsider this, fitting only the data at $z\leq1.0$ ($z\leq0.5$), and find the same trend 
at $\sim6\sigma$ ($\sim4\sigma$). 

%\clearpage
%\clearpage

\begin{figure*}
    \centering
    \figexpand
    \plotone{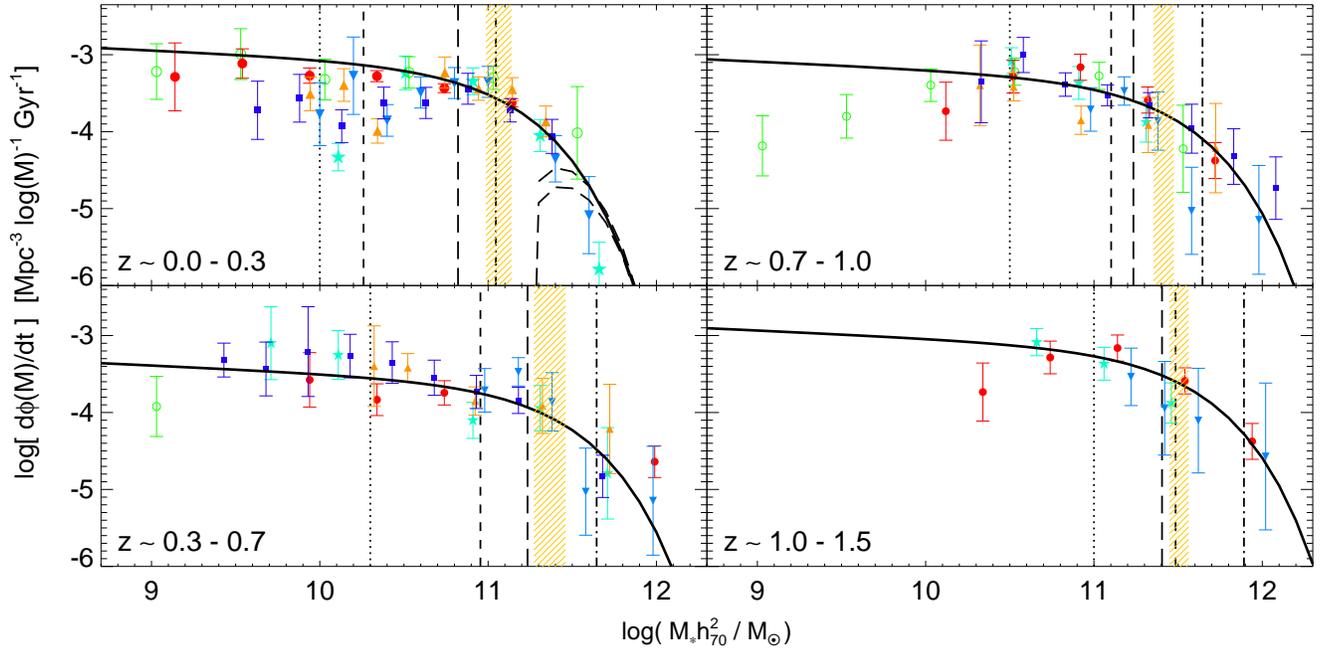}
    \caption{Time-averaged rate of formation of elliptical 
    galaxies, obtained by differentiating the observed elliptical mass 
    functions in Figure~\ref{fig:oplot.mfs} with respect to time 
    (colored points; style as in Figure~\ref{fig:compare.methods}).
    Results are shown over a number of redshift intervals, as labeled. 
    Dotted lines show typical 
    completeness limits at each redshift. Short-dashed, long-dashed, and 
    dash-dotted vertical lines 
    show $M_{50}$, $\mtrans$, and $M_{Q}$ 
    from Figure~\ref{fig:compare.methods} (color-selected fits) at each $z$.
    Solid line in each panel shows the best-fit Schechter function 
    (to points above the completeness limits quoted for each sample), 
    with the shaded range showing the $1\sigma$ range of the best-fit 
    $M_{\ast}$. High mass ellipticals appear to preferentially build up 
    at higher redshifts. 
    The characteristic masses in Figure~\ref{fig:compare.methods}
    provides a reasonable proxy for the characteristic $M_{\ast}$ being added or 
    ``built up'' in the early-type galaxy mass function at each redshift, 
    as both evolve to higher masses at higher redshift. 
    In the upper panel, thick dashed lines 
    show the maximal contribution from spheroid-spheroid mergers 
    if all undergo $\sim0.5-1$ (lower and upper lines) such mergers 
    in that redshift interval (at or below the break at each $z$, a significant 
    dissipationless merging fraction will {\em lower} the number density 
    at low $M$, where it is observed to rise). 
    \label{fig:ell.buildup.mfs}}
    %\epsscale{1.0}
\end{figure*}

Figure~\ref{fig:ell.buildup.mfs} considers the differential growth of the early-type 
MF in more detail. We show the time-averaged buildup of early-type MFs in several 
redshift intervals from $z=0-1.5$. Where a given sample measures the elliptical mass function at 
two redshifts $z$ and $z+\Delta z$, we differentiate 
the observed elliptical mass function at every mass with respect to the 
two redshifts to obtain the time-averaged rate of creation of ellipticals of that 
mass, over that redshift interval. In total, we 
show four redshift intervals: $z\sim0.0-0.3$, $0.3-0.7$, $0.7-1.0$, and $1.0-1.5$. Of course, we 
only compare mass functions measured with the same technique and sample in the same study, 
since systematic offsets in methodology could severely bias 
such an estimate. 
Although the scatter is large (especially at low 
masses), the observations all trace a similar elliptical formation rate as a function 
of mass, with a similar break traced 
in the different samples at each redshift. The shape of this function is not
the same as that of the early type mass functions -- i.e.\ 
we are not simply recovering the fact that the mass function builds up uniformly 
over these intervals.

%\clearpage
%\clearpage

At low redshift ($z\lesssim0.3$), comparison with the early-type MFs in Figure~\ref{fig:oplot.mfs} shows 
that sub-$M_{\ast}$ ($M\lesssim10^{11}\,M_{\sun}$) ellipticals are ``building up'' 
in number density by $\sim7-15\%$ per Gyr, whereas the most massive 
systems build up by only $\sim1\%$ per Gyr. In other words, the most 
massive systems are not building up (either via star formation {\em or} assembly of 
stellar populations) at a significant rate at low redshift. 
At high $z$, there is a marginal shift of this function to higher masses. By $z\sim1$, 
comparison with the corresponding early-type MFs implies that systems 
with $M_{\ast}\gtrsim10^{11}\,M_{\sun}$ are building up by $\sim20-50\%$ per Gyr. 
The best-fit Schechter functions plotted in each redshift interval reflect this, 
with the Schechter $M_{\ast}$ shifting from $11.02\pm0.11$ ($z\lesssim0.3$) 
to $11.51\pm0.07$ ($1.0\lesssim z \lesssim 1.5$). 

The low-$z$ growth estimate of $\sim1\%$ per Gyr in the most massive systems
is in excellent agreement with that from \citet{Masjedi06} and \citet{Bell06b} determined from local 
red galaxy (spheroid-spheroid or ``dry'') merger rates. 
In detail, observations suggest that the 
typical massive red galaxy undergoes $\sim0.5-1$ major dissipationless mergers since 
$z=1$ \citep{vanDokkum05,Bell06}. If this is representative, it is trivial to predict the corresponding 
rate of buildup of the elliptical population, assuming every red galaxy undergoes 
this number of major ($\sim1:1$ mass ratio) 
mergers in this time. This is shown in Figure~\ref{fig:ell.buildup.mfs}, calculated from 
the local MF of \citet{Bell03}. Although these mergers appear to be important
for building up the most massive galaxies at low ($z\lesssim0.3$) redshift, 
their contribution cuts off completely below $\sim2\times10^{11}\,M_{\sun}$ (and will cut off at higher 
masses at higher $z$, generally about $\sim2$ times the break in the ``buildup'' mass 
function). Such mergers move galaxies 
from the low-mass end of the elliptical mass function to the high mass end, and since the 
low-mass slope of the elliptical mass function is not steep, this can only 
{\em decrease} the number density of low-mass objects. 
By definition, then, dry mergers cannot account for the (substantial) buildup of 
{\em total} mass on the RS nor the buildup at low and intermediate masses.
If these lower-mass ellipticals 
are formed by gas-rich mergers, 
then there must be at least $\sim2$ times as many 
gas-rich mergers moving new galaxies onto the elliptical mass function 
as there are dry mergers (in fact, the actual observed ratio is $\sim10:1$, \citet{Bell06}).

\section{The Transition Mass and Characteristic Galaxy Masses}
\label{sec:compare.mstar}

\begin{figure*}
    \centering
    %\epsscale{1.0}
    \plotone{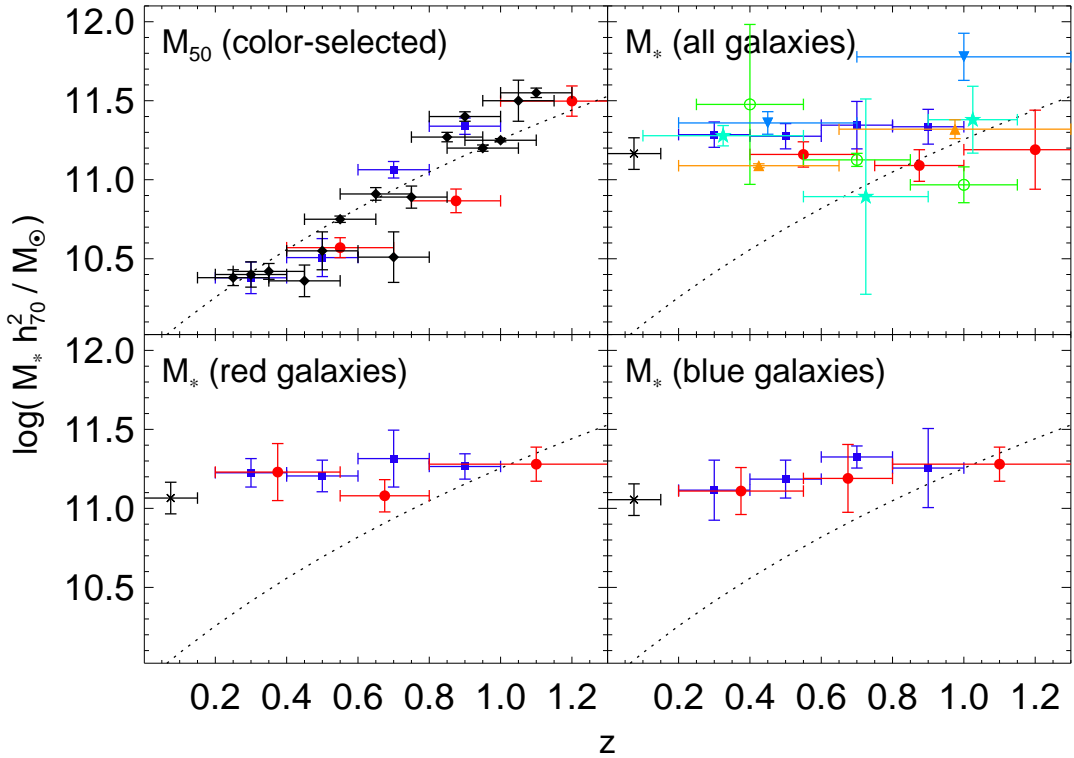}
    \caption{Comparison of the ``downsizing'' or ``transition mass'' with 
    characteristic galaxy masses. 
    Upper left shows $M_{50}$ (points; from color-selected samples) and 
    best-fit trend (dotted line) from Figure~\ref{fig:compare.methods}. 
    Other panels show the best-fit Schechter function $M_{\ast}$ to 
    the all-galaxy mass functions, and color-selected red and blue galaxy 
    mass functions (points as in Figure~\ref{fig:compare.methods}). 
    The trend in 
    $M_{50}$ (and $M_{tr}$, $M_{Q}$) does not trace the all, blue, or 
    red galaxy populations -- i.e.\ ``transition mass'' systems are not 
    uniformly drawn from any of these populations. Note that the trend 
    in $M_{50}$ is not incompatible with that in $M_{\ast}$ of red galaxies, as 
    the Schechter function parameterization is such that an order-of-magnitude 
    change in $M_{50}$ at a constant rate of elliptical ``formation'' implies only a 
    $\sim0.1-0.2$ dex change in $M_{\ast}$ (which can be further offset by a small 
    dry merger rate). 
    \label{fig:compare.mcrit.mstar}}
    %\epsscale{1.0}
\end{figure*}

Figure~\ref{fig:compare.mcrit.mstar} compares the ``transition'' 
mass determined above with the characteristic masses (Schechter $M_{\ast}$) 
of the red, blue, and all galaxy populations. For clarity, we show just
$M_{50}$, as it is the most well-determined of the masses in Figure~\ref{fig:compare.methods}, 
as well as being most robust with respect to sample definitions/selection, 
and further it has the most direct physical interpretation (as it is not degenerate 
with blue cloud evolution). Our conclusions here and subsequently, however, are 
unchanged regardless of the mass definition from Figure~\ref{fig:compare.methods}. 
The characteristic masses of red, blue, and all galaxies are nearly constant with 
redshift, with at most a marginal ($\sim0.2$\,dex) increase from $z=0-1$; inconsistent 
with their following the strong trend seen in $M_{50}$ at $\gtrsim10\sigma$ (independent of 
normalization). Likewise, comparing the shape of the 
rate of elliptical buildup in Figure~\ref{fig:ell.buildup.mfs} with these galaxy mass 
functions at the same redshift shows that they do not trace the same mass 
distribution as a function of redshift. We can therefore (perhaps unsurprisingly, given 
the definitions employed)
rule out at high significance 
the hypothesis that ``transition'' mass objects are uniformly/randomly 
drawn from a ``parent'' population of normal galaxies of either early or late (or both) types.

%\clearpage
%\clearpage

It may appear that the strong trend in $M_{50}$ (i.e.\ the mass above which 
the RS MF is $>50\%$ assembled at $z$) is incompatible with the weak trend 
in $M_{\ast}$ of red galaxies (as e.g.\ no change in $M_{\ast}$ would imply 
uniform buildup of RS populations at all masses). This, however, is 
an artifact of the Schechter function fit. For example, given a local 
early-type Schechter function MF with $\phi_{\ast}=\phi_{0}$, 
$\alpha$, and $M_{\ast}=M_{0}$ and a similar MF at $z$ with 
$\phi_{\ast}=\phi_{z}$, the same $\alpha$ (commonly assumed in fitting), 
and $M_{\ast}=M_{0}\,(1+\delta)$, one obtains 
\begin{equation}
M_{50}(z) = M_{0}\,{\Bigl(}{1-\frac{1}{1+\delta}}{\Bigr)}^{-1}\,
\ln{{\Bigl[}{\frac{\phi_{0}}{2\,\phi_{z}}(1+\delta)^{\alpha+1}}{\Bigr]}}, 
\end{equation}
which is quite sensitive to $\delta$ and, for the observed values of 
$\phi_{\ast}(z)$, $M_{\ast}(z)$ \citep[e.g.,][]{Borch06} 
predicts a $\sim1-2$\,dex evolution in 
$M_{50}$ similar to what we find. 

In a more physical sense, the 
local RS MF is the sum of the RS MFs built up over various intervals, 
each of which resembles a Schechter function (see Figure~\ref{fig:ell.buildup.mfs}). 
Consider the sum of two Schechter functions with identical $\alpha$ and $\phi_{\ast}$, 
but one having lower $M_{\ast}$ by $1\,$dex. Fitting this to a Schechter function 
over the range $M_{\ast}\pm1$\,dex ($\pm1.5\,$dex) yields a 
best-fit with $M_{\ast}$ only lowered from the higher value by $\sim0.1\,$dex ($0.05\,$dex). 
This owes to the steep fall in $\phi(M)$ at $M\gg M_{\ast}$, which forces the combined 
fit to retain the high $M_{\ast}$ value. In further detail, if we imagine the 
rate of buildup ${\rm d}\phi(M)/{\rm d}t(z)$ 
(Figure~\ref{fig:ell.buildup.mfs}) is a Schechter function 
with constant $\alpha$ and normalization, but an evolving $M_{\ast}=M_{50}(z)$ 
(adopting the best-fit trend $M_{50}(z)$ shown in Figure~\ref{fig:compare.mcrit.mstar}), 
then fit the integrated $\phi_{\rm early}(M,\,z)$ to a Schechter function (fitting 
over the range $M_{\ast}\pm1$\,dex), we obtain only $\sim0.15-0.20$\,dex 
evolution in the early-type $M_{\ast}$ from $z=0$ to $z=1$, despite the 
more than order-of-magnitude evolution in $M_{50}(z)$. 
Thus, although the strong evolution in the ``transition'' mass with redshift 
rules out its being representative of the general elliptical population, 
it is {\em not} inconsistent with the weak evolution in the early-type $M_{\ast}$, 
even if $M_{50}$ does represent the characteristic 
mass at which ``new'' galaxies are being added to the red sequence. In other 
words, weak evolution in $M_{\ast}$ of red galaxies does not rule out 
strong evolution in the characteristic masses being ``built up'' on or added to 
the RS.

\section{The Transition Mass and Mergers}
\label{sec:compare.merger.mass}

We next consider observed merger MFs. 
We compile the local ($z\leq0.2$) pair-selected major (within $\sim1$\,mag) merger 
luminosity functions from \citet[][2MASS]{Xu04} in 
$K$-band and \citet[][]{Toledo99} in $B$-band, as well 
as the morphologically identified merger/interacting galaxy luminosity 
functions from \citet[][CFRS+LDSS; $z\leq1$]{Brinchmann98} in $B$-band 
and \citet[][GEMS+GOODS; $z\sim0.7$]{Wolf05} in the near-UV ($280\,{\rm nm}$), 
and mass functions from 
\citet[][HDF-N and HDF-S; $z\sim1-3$]{Conselice03,Conselice05}
and \citet[][GOODS+DEEP2; $z\leq1.4$]{Bundy05.masslimit}. 
Where the MFs are not directly measured, 
we rescale the luminosity functions to mass functions using the 
mass-to-light ratios of ongoing mergers (as a function of $M_{\ast}$) 
from \citet[][see Table~1]{H06d}. These are calculated from 
the population synthesis models of \citet{BC03}, given the distribution of 
star formation histories 
during mergers determined from several hundred 
numerical simulations 
that include star formation, supernova feedback and metal 
enrichment, and black hole accretion and feedback (see Hopkins et al.\ 2006d for details; 
this is essentially a second-order improvement on the typical empirically adopted simplified 
tau+burst models for these $M/L$). These should 
be reasonably robust: they have 
also been checked directly in the bands of interest here 
against the measurements of $M/L$ ratios in 
local ULIRGs \citep{Tacconi02}, pair samples \citep{Dasyra06}, and 
recent merger remnants \citep{RJ04}, and give good agreement 
\citep[][Figures~1 \&\ 4]{H06d}.  Furthermore,
\citet[][Figures~8 \&\ 9]{H06d} demonstrate that they can be reliably used to 
convert merger luminosity functions to mass functions (in exactly this manner) for 
all samples above where both are measured. In any case, our subsequent results 
are unchanged (albeit their significance reduced given the limited data) 
if we consider only the morphologically identified, 
directly measured merger mass functions of 
\citet{Conselice03,Conselice05} and \citet{Bundy05.masslimit}. 

\begin{figure*}
    \centering
    \figexpand
    \plotone{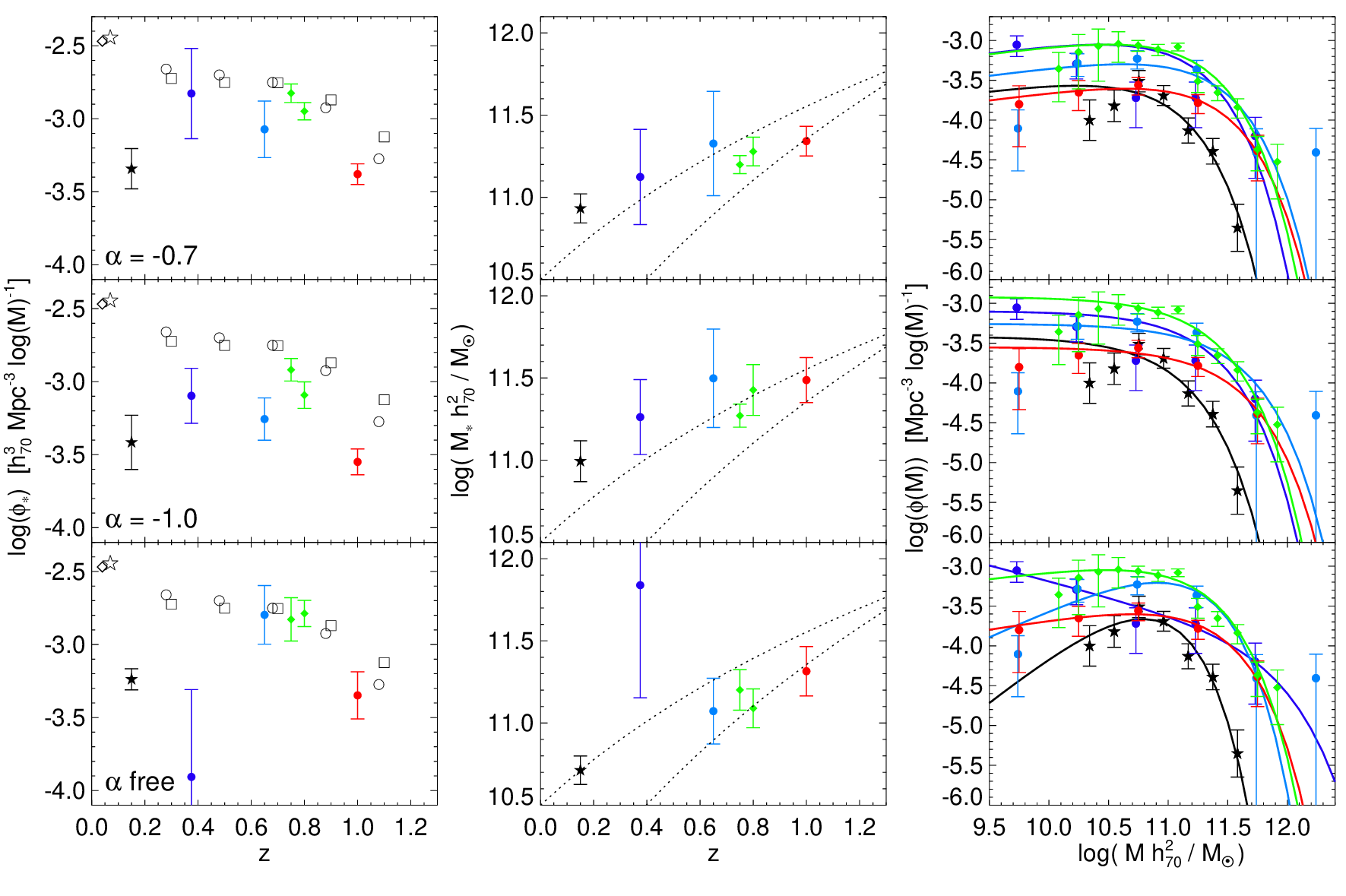}
    \caption{Comparison of observed merger mass functions from \citet[][black stars]{Xu04} 
    at $z\lesssim0.2$, 
    \citet[][circles]{Bundy05.full,Bundy05.masslimit} at $z=0.2-0.5$ (purple), 
    $z=0.5-0.8$ (blue), and $z=0.8-1.2$ (red), and \citet[][green diamonds]{Wolf05} at 
    $z=0.7-1.0$ (upper $\&$ lower $z$ points from GEMS and GOODS, respectively). 
    Panels show the merger mass functions (right), with best-fit Schechter functions (lines of 
    corresponding color), and best-fit Schechter function $\phi_{\ast}$ (left) and 
    $M_{\ast}$ (center). Because constraints on the faint-end slope are weak, 
    upper panels fix its value $\alpha=-0.7$, middle 
    panels fix $\alpha=-1.0$, and lower panels allow $\alpha$ to be fit. 
    Open points in the left panels show $\phi_{\ast}$ of the entire galaxy population 
    as compiled in \citet{Faber05} 
    from \citet[][SDSS; star]{Bell03}, \citet[][2dF; diamond]{Madgwick03}, 
    \citet[][COMBO-17; circles]{Bell04}, and \citet[][DEEP2; squares]{Willmer05}. 
    Dotted lines in center panels show the fitted 
    $M_{50}(z)$ (lower) and $\mtrans(z)$ from Figure~\ref{fig:compare.methods} (color-selected). 
    Regardless of the choice of $\alpha$, there is a trend for the characteristic merger 
    mass $M_{\ast}$ to increase with redshift in a manner similar to $\mtrans$. 
    Interpretation of $\phi_{\ast}$ is more ambiguous, but there is a suggestion that it 
    increases {\em relative} to the $\phi_{\ast}$ of the galaxy population. 
    \label{fig:merger.mfs}}
    %\epsscale{1.0}
\end{figure*}

Figure~\ref{fig:merger.mfs} shows several (the most well-constrained) of these 
MFs as a function of redshift, with the best-fit Schechter functions. Since 
constraints are weak at the faint end (and systematic uncertainties large; discussed below), 
we consider fits with fixed faint-end slopes $\alpha$, as well as 
allowing $\alpha$ to vary. In all cases, there is a qualitatively similar 
trend for the Schechter function $M_{\ast}$ to increase (by $\sim0.5-0.7$\,dex from $z=0-1$). 
Direct comparison of the MFs demonstrates that this does not necessarily mean 
the number of mergers at the high-mass end increases monotonically with redshift (changes 
in $\phi_{\ast}$ can offset the increase in $M_{\ast}$); 
this does however imply that the {\em relative} merger history/MFs 
favors higher mass mergers at higher redshifts. Although it is not important for our 
comparison, there is also a marginal trend for $\phi_{\ast}$ of mergers to increase 
with redshift relative to 
$\phi_{\ast}$ of the entire galaxy population, but this trend is largely driven by the 
low $\phi_{\ast}$ from \citet{Xu04}.

%\clearpage
%\clearpage

\begin{figure*}
    \centering
    \figexpand
    \plotone{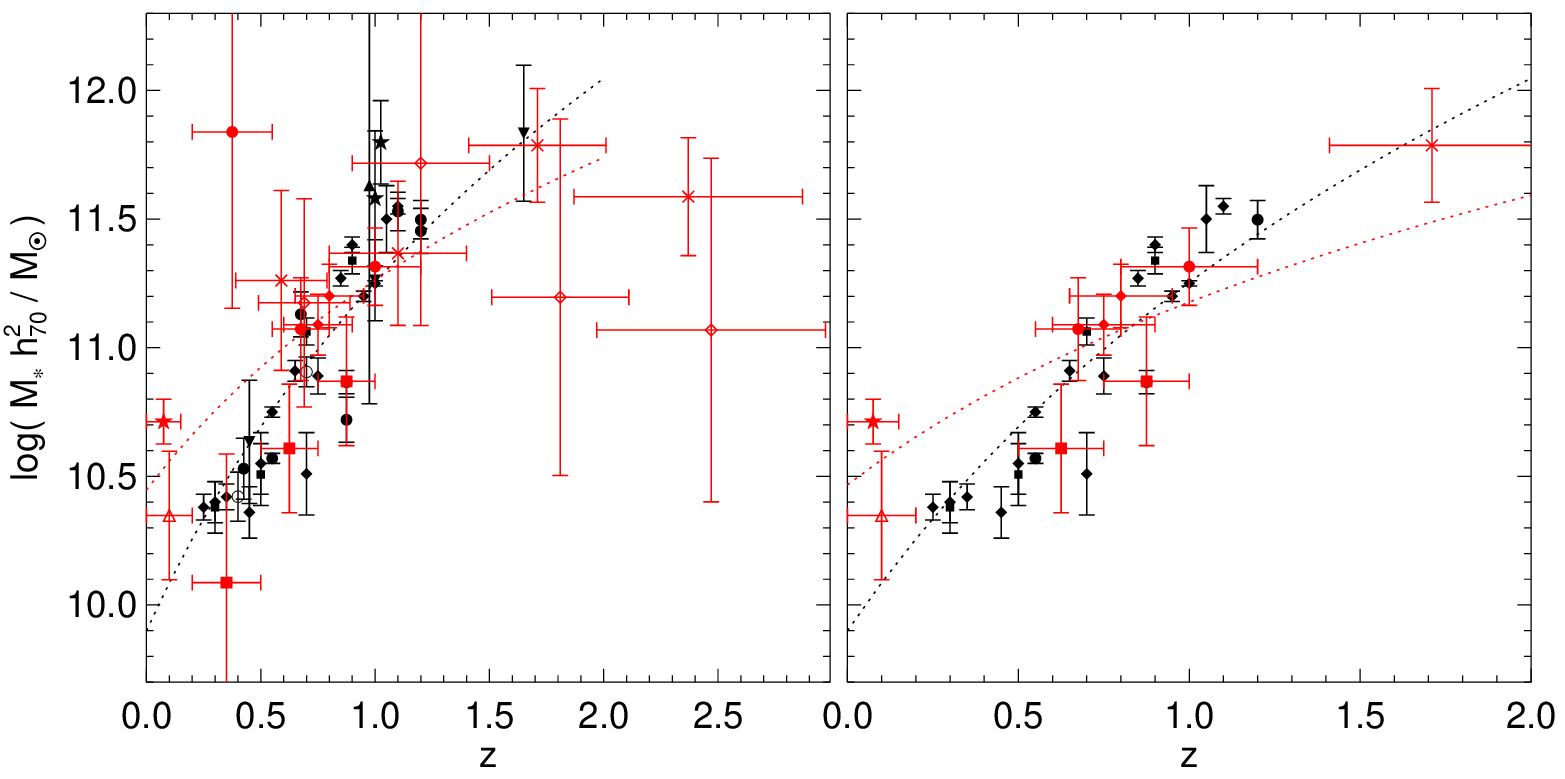}
    \caption{{\em Left:} ``Transition'' mass $M_{50}$ from all sample selections
    shown in Figure~\ref{fig:compare.methods} (black points in same style), 
    and the best-fit trend with redshift (black dotted line), compared to the 
    characteristic mass $M_{\ast}$ from merger mass and luminosity 
    functions (red points; see Figure~\ref{fig:merger.mfs}), from 
    \citet[][stars]{Xu04}, \citet[][filled diamonds]{Wolf05}, 
    \citet[][circles]{Bundy05.full,Bundy05.masslimit}, 
    \citet[][HDF-S, crosses; HDF-N, open diamonds]{Conselice03}, \citet[][triangles]{Toledo99}, 
    and \citet[][squares]{Brinchmann98}, 
    with the best-fit trend of the form $M_{\ast}\propto(1+z)^{\kappa}$ (red dotted line). 
    The $M_{\ast}$ values shown allow the merger 
    mass function faint-end slope $\alpha$ to vary freely, but a similar result is obtained 
    fixing $\alpha$ to match the early-type or all-galaxy values (see Figure~\ref{fig:merger.mfs}). 
    {\em Right:} Same, but for clarity, only $M_{50}$ from color-selected samples 
    and the best-constrained $1/2$ of merger $M_{\ast}$ values are shown. Despite 
    the small sample sizes, the characteristic mass of merger mass functions 
    increases with redshift at $>3\sigma$ (implying mergers are not simply drawn from 
    the approximately constant $M_{\ast}$ all, blue, or red galaxy populations) and
    is consistent with the value and evolution of $M_{50}$ as a function of redshift, 
    as in Figure~\ref{fig:ell.buildup.vs.merger.mfs}.  
    \label{fig:compare.mergers}}
    %\epsscale{1.0}
\end{figure*}

Figure~\ref{fig:compare.mergers} plots these Schechter $M_{\ast}$ values, 
as well as those determined from the other merger mass/luminosity functions we compile, as a 
function of redshift, compared to the characteristic ``transition'' mass ($M_{50}$). 
The characteristic masses from pair and morphologically-selected samples, as 
well as direct MFs, optical, near-IR, and near-UV luminosity functions agree 
surprisingly well at overlapping redshifts, at least up to the $z\gtrsim2$ estimates which are 
strongly affected by cosmic variance \citep{Conselice05}, which further suggests the 
typical merger $M/L$ ratios used are reasonable. There is a significant ($>3\sigma$) 
trend for the characteristic masses of mergers to increase with redshift. As with 
the ``transition'' mass population in Figure~\ref{fig:compare.mcrit.mstar}, 
this trend rules out at $>3\sigma$ the hypothesis that mergers are randomly/uniformly 
drawn from the all or red galaxy population. Whether mergers uniformly trace the 
blue galaxy population is less clear; the values plotted are inconsistent with this 
hypothesis at $\sim3.5\sigma$, but the trend alone (i.e.\ allowing for a 
systematic normalization offset) is inconsistent at only $\sim2\sigma$. The values/trend of  
the merger MF $M_{\ast}$ as a function of redshift are, 
however, similar and statistically consistent with $M_{50}$ 
(and even more similar to $\mtrans$ and $M_{Q}$, see Figure~\ref{fig:compare.models}). 

%\clearpage
%\clearpage

\begin{figure*}
    \centering
    \figexpand
    %\epsscale{1.1}
    \plotone{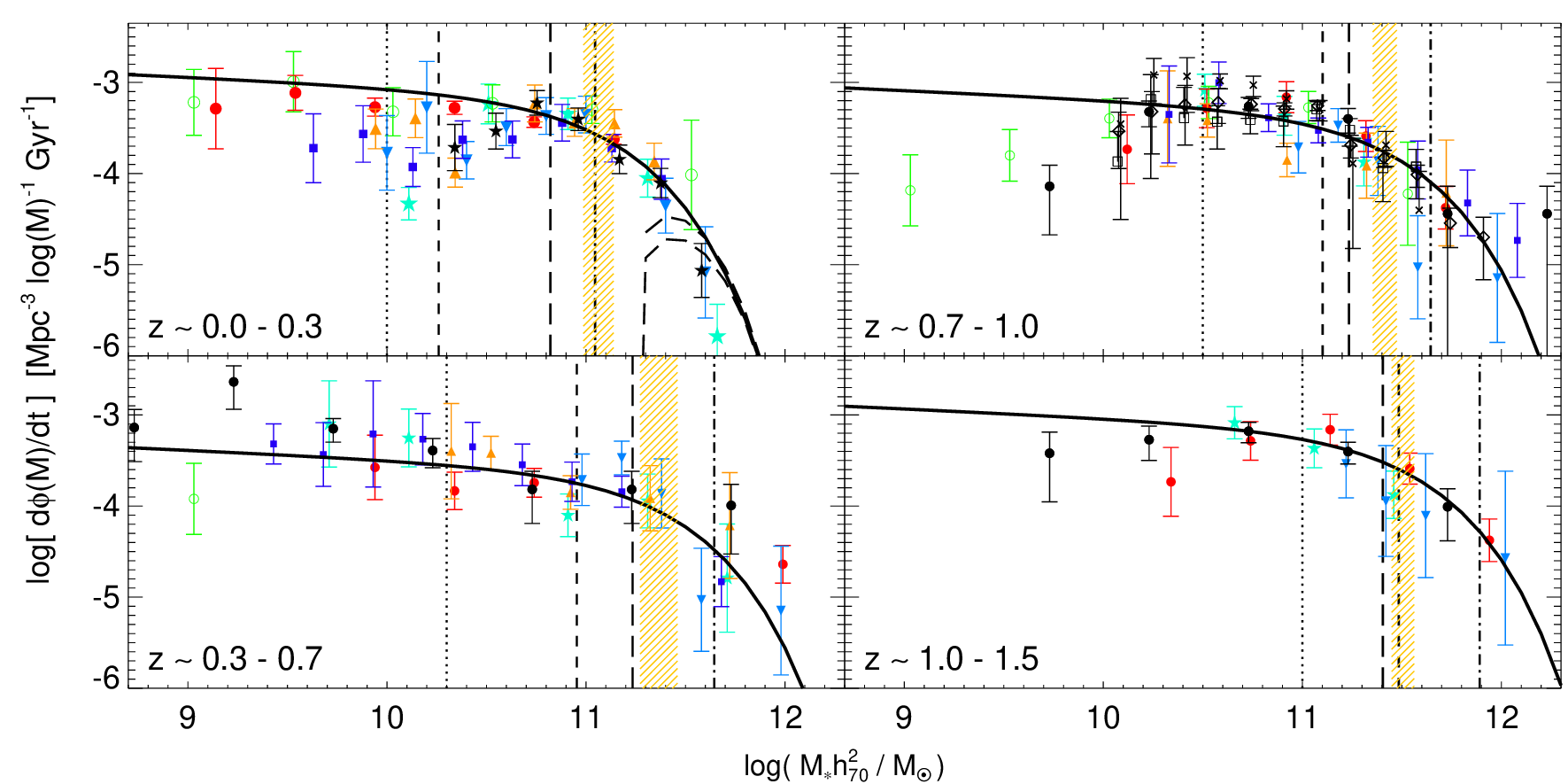}
    \caption{Time-averaged rate of buildup of early-type populations, as in 
    Figure~\ref{fig:ell.buildup.mfs}, but with the observed merger mass functions 
    from Figure~\ref{fig:merger.mfs} overlaid (black points, style as in Figure~\ref{fig:merger.mfs}). 
    Merger mass function 
    points at $z\sim0.7-1.0$ from \citet{Wolf05} show the effects of systematic differences in 
    imaging depth and survey area, from GEMS (squares), GOODS ($\times$'s), 
    and GEMS+GOODS (diamonds).
    Since the merger {\em timescale} is observationally 
    undetermined, we compare the mass {\em distribution} in mergers and 
    the early-type buildup (i.e.\ merger mass function are renormalized for direct comparison). 
    The time-averaged buildup of elliptical populations traces a similar mass distribution to 
    that of observed merger populations at all redshifts observed, and not the same mass 
    distribution as that of the all, red, or blue galaxy population as a function of redshift 
    (compare Figure~\ref{fig:compare.mcrit.mstar}). 
    \label{fig:ell.buildup.vs.merger.mfs}}
    %\epsscale{1.0}
\end{figure*}

We can consider in greater detail if observed merger mass/luminosity functions 
are consistent with the observed buildup of early-type populations by 
examining 
the complete mass functions as a function of redshift. 
Figure~\ref{fig:ell.buildup.vs.merger.mfs} reproduces Figure~\ref{fig:ell.buildup.mfs}, 
but overlays the observed merger MFs from Figure~\ref{fig:merger.mfs} 
at the appropriate observed redshifts. Since 
there is considerable ambiguity in converting an observed merger 
MF to a merger {\em rate}, we renormalize (vertically only; i.e.\ divide out an 
appropriate timescale) the observed MFs arbitrarily 
such that we can focus here just on the unambiguous mass distribution (although 
we will consider the issue of absolute rates/normalization shortly). The agreement 
is striking: independent of the systematics in understanding merger rates, 
the observed distribution of mergers as a function of mass and redshift traces and is 
consistent with the buildup/addition of galaxies to the red sequence at all masses 
and redshifts observed.

%\clearpage
%\clearpage

Although systematically uncertain, we should also compare the implied 
merger rates (i.e.\ vertical normalization in Figure~\ref{fig:ell.buildup.vs.merger.mfs}). 
In other words, even if the buildup 
of early-type populations traces the merger mass distribution, are there 
the appropriate total number/rate of mergers to account for the growth 
of the RS MF (assuming mergers are the agent of this ``buildup'')? To estimate this, 
we assume every elliptical formation/addition event in Figure~\ref{fig:ell.buildup.mfs} 
indeed owes to a merger, which is observable as such for 
some amount of time $t_{\rm merger}$
(until morphological disturbances such as tidal tails fade beyond 
typical surface brightness limits). This yields the expected 
merger MF, $\phi(M_{\rm merger})\sim t_{\rm merger}\,\dot{\phi}(M_{\rm gal})$. 
Given the galaxy MFs from which we calculated these rates in the 
first place, it is then trivial to estimate the expected 
merger fraction as a function of mass. 

\begin{figure*}
    \centering
    \figexpand
    \plotone{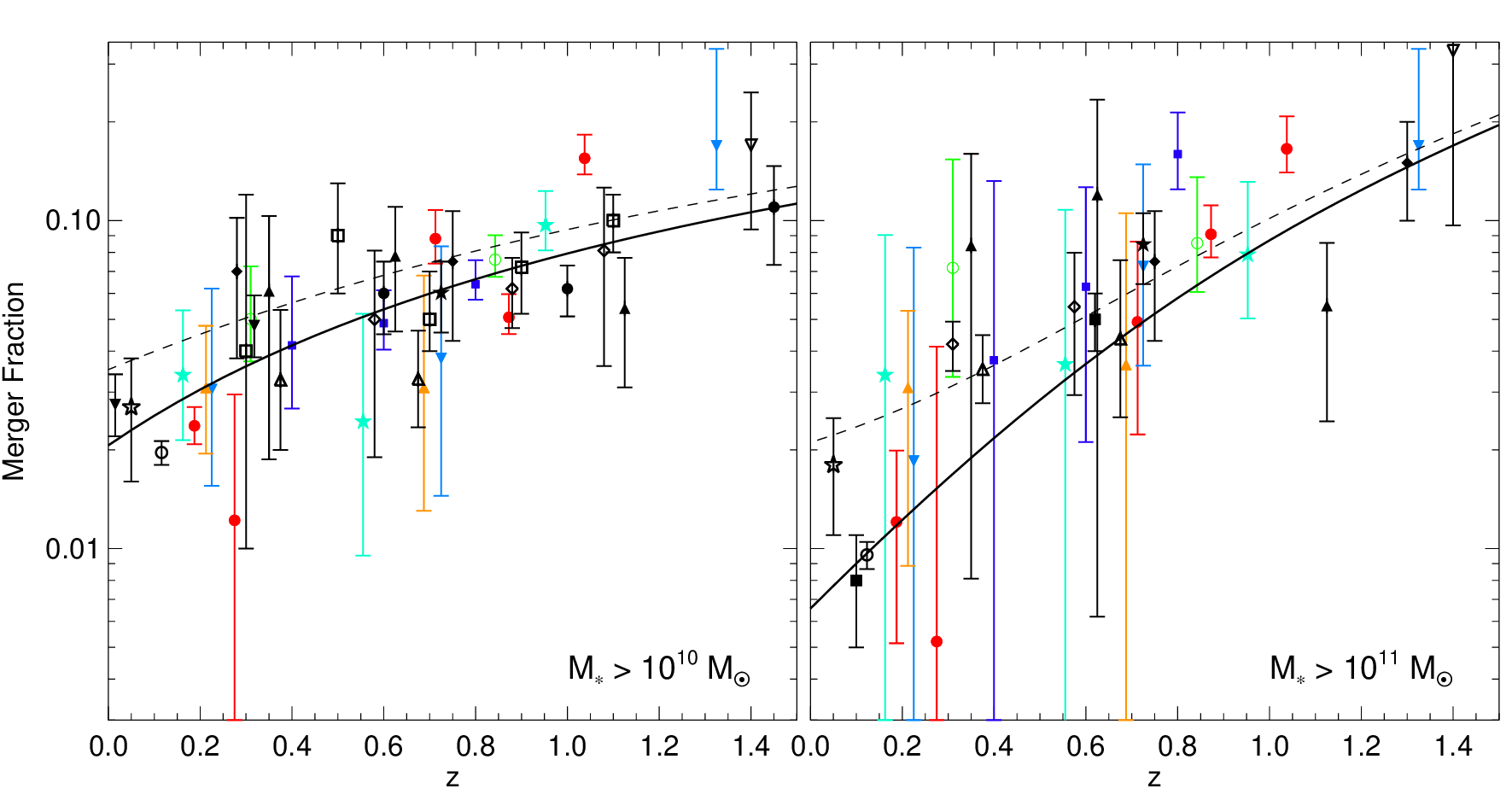}
    \caption{Expected merger fraction as a function of redshift, if the buildup of the 
    early-type mass functions in Figure~\ref{fig:merger.mfs} 
    is entirely a result of mergers moving galaxies onto 
    the red sequence (colored points as in Figure~\ref{fig:merger.mfs}). 
    Black points show observed merger fractions as a function of redshift, 
    from  \citet[][filled inverted triangles]{Patton02}, \citet[][filled circles]{Conselice03}, 
    \citet[][filled triangles]{Bundy04}, 
    \citet[][open diamonds]{Lin04}, \citet[][open stars]{Xu04}, 
    \citet[][open circles]{dePropris05}, \citet[][filled diamonds]{Cassata05}, 
    \citet[][filled stars]{Wolf05}, \citet[][open triangles]{Bundy05.full},     
    \citet[][open inverted triangles]{Lotz06a}, \citet[][open squares]{Lotz06b},
    and \citet[][filled squares]{Bell06b}, 
    Results are shown for two (approximate) 
    minimum stellar mass limits, as labeled. 
    Solid line shows the expected gas-rich merger fraction 
    if all bright quasars are triggered in mergers, using the same modeling 
    from Figure~\ref{fig:compare.quasars}
    to determine the quasar-parent mass function from the observed 
    quasar luminosity function. 
    Dashed line adds a constant fraction 
    \citep[observed $0.015$;][]{Bell06,Lotz06b} 
    of dissipationless (spheroid-spheroid) mergers.
    An observable merger timescale of $0.5\,$Gyr is assumed. 
    The expected merger fractions from the observed buildup of early-type mass 
    functions and the quasar luminosity function agree reasonably in their normalization 
    and evolution with observed merger fractions. There are sufficient mergers 
    to account for both populations, and little room for a large fraction of mergers 
    which do not produce a remnant elliptical or trigger quasar activity. Dissipationless 
    mergers are generally a relatively small effect 
    as is observed, but may be important for the buildup of the 
    most massive systems at low ($z\lesssim0.5$) redshifts. 
    \label{fig:merger.frac}}
    %\epsscale{1.0}
\end{figure*}

Figure~\ref{fig:merger.frac} compares this estimate, adopting a 
characteristic $t_{\rm merger}=0.5\,{\rm Gyr}$, with observed 
merger fractions as a function of mass and redshift. This 
timescale is roughly expected from numerical simulations of mergers 
\citep[e.g.,][]{Robertson05a}, dynamical friction considerations \citep{Patton02}, 
or more detailed estimates of observational selection effects 
as a function of merger stage \citep[see, e.g.,][for a detailed discussion of these 
issues]{H06d}. We consider 
the merger fraction above two representative stellar mass limits, 
$M_{\ast}>10^{10}\,M_{\sun}$ and $M_{\ast}>10^{11}\,M_{\sun}$, but caution that these 
are not, for the most part, mass-selected samples, so the mass limits shown
in Figure~\ref{fig:merger.frac} are only broadly applicable. 
We calculate the merger fraction as a function 
of mass directly from the merger mass functions of \citet{Xu04,Wolf05,Bundy05.full} 
shown in Figure~\ref{fig:merger.mfs}, at redshifts where the samples 
are complete to the given mass limit (this essentially excludes the $z\gtrsim1$ 
\citet{Bundy05.full} merger mass function). 

%\clearpage
%\clearpage

The observed merger fractions are consistent with this estimate at all redshifts. 
The buildup of elliptical populations does suggest that merger fractions should 
increase as a function of redshift, but we note that the effect is quite weak. 
If, for example, the characteristic merger timescale decreases with redshift in the same manner 
as halo dynamical times (at fixed mass), $\propto(1+z)^{-3/2}$ (decreasing the 
expected merger fraction we calculate by this amount), the expected 
increase in merger fraction with redshift becomes marginal (factor $\lesssim2$ by 
$z\sim1.5$). Both cases, however, are consistent with the present observations 
(at these redshifts, within the factor $\sim2$ typical uncertainties). 
There is also marginal evidence for steeper evolution in merger fractions with 
redshift in the higher-mass cut we consider, evidence for which is 
also seen in e.g.\ \citet{Conselice03,Conselice05}, but we caution both that 
this trend appears only when different samples are combined, and that it 
will be ``washed out'' by the increasing importance of dry mergers at high 
masses and low redshifts (see also Figure~\ref{fig:ell.buildup.vs.qso}). 
Future studies which can separate gas-rich and gas-poor merger  
populations and track the merger fraction as a function of redshift 
and mass can provide a substantially stronger test of these trends. 

Caution regarding systematic uncertainties in merger populations is 
still warranted. \citet{LeFevre00} find that pair and morphological
selection criteria yield similar results, but \citet{Lin04} see
significant disagreement, which may be the result of systematic identification of ``false'' 
(i.e.\ non-merging) pairs 
at low redshift \citep{Berrier06}. Fortunately, the characteristic merger mass 
(or luminosity) does not appear to change
dramatically with selection method even though the time spent in a
given phase (and thus $\phi_{\ast}$ or merger fraction) may. For example,
the data of \citet{Lin04} and \citet{Conselice03} do yield a similar
characteristic merging galaxy luminosity $L_{\ast}$, 
despite finding different merger fractions. 
One might also wonder
whether the natural tendency of a flux-limited sample to select 
brighter systems at higher redshift might lead one to infer an increasing 
mass scale regardless of the underlying mass distribution. However, 
Figure~\ref{fig:merger.mfs} demonstrates that 
the completeness limits for most of the samples we consider 
are generally well below the ``transition'' mass, and similarly below 
the break in the corresponding mass function. 
Further, although
\citet{Wolf05} find that the observed number of faint
mergers depends on selection effects (see Figure~\ref{fig:ell.buildup.vs.merger.mfs}; 
which shows the increase in number of low-mass mergers when 
increasing imaging depth and decreasing field size by 
an order of magnitude, from GEMS to GOODS), this does not 
significantly effect the merger $M_{\ast}$ or change our 
comparison in Figure~\ref{fig:ell.buildup.vs.merger.mfs}. Still, these effects must be accounted for 
in any comparison of fitted MFs. 

It is furthermore true that the exact 
appropriate value of the duration of observable merger activity ($t_{\rm merger}$) 
is not well-determined, and will in detail depend on the sample, mass limit, 
and redshift, but for our purposes these effects (amounting to a 
systematic factor $\sim2$ uncertainty) are generally comparable 
to or smaller than the scatter in the observations in Figure~\ref{fig:merger.frac}. 
Preliminary estimates of the observable merger timescale 
based on comparison with 
automated nonparametric classification schemes \citep{Lotz04} suggest, 
perhaps surprisingly, relatively weak trends with redshift 
(at least at $z\lesssim2$; see the discussion in Hopkins et al.\ 2006d), but lacking 
a complete cosmological framework from which to predict observable 
merger properties, Figure~\ref{fig:merger.frac} should be taken with the strong caveat 
that the relative normalizations of galaxy buildup and observed mergers 
depends systematically on $t_{\rm merger}$.

\section{The Transition Mass and E+A Galaxies}
\label{sec:ea}

Detailed studies of ``E+A'' (or ``K+A'') 
galaxies \citep{DresslerGunn83}, with characteristic 
post-starburst stellar populations indicating a substantial but rapidly 
quenched star formation epoch in the last $\sim0.1-0.5$\,Gyr 
\citep{Caldwell96,CS87,Quintero04}, have found ubiquitous 
evidence of morphological disturbances and tidal tails 
\citep{Schweizer96,Blake04,Goto05}, which together with their 
environmental \citep[e.g.,][]{Zabludoff96,Goto05} and structural/kinematic 
properties \citep{Kelson00,Norton01,Tran03,vanderWel04} imply their 
formation in mergers and evolution into typical early-type galaxies. We therefore 
consider whether E+A galaxies, presumably recently formed in mergers, 
trace any ``downsizing'' trend. 

%\clearpage
\begin{figure}
    \centering
    \figexpand
    \plotone{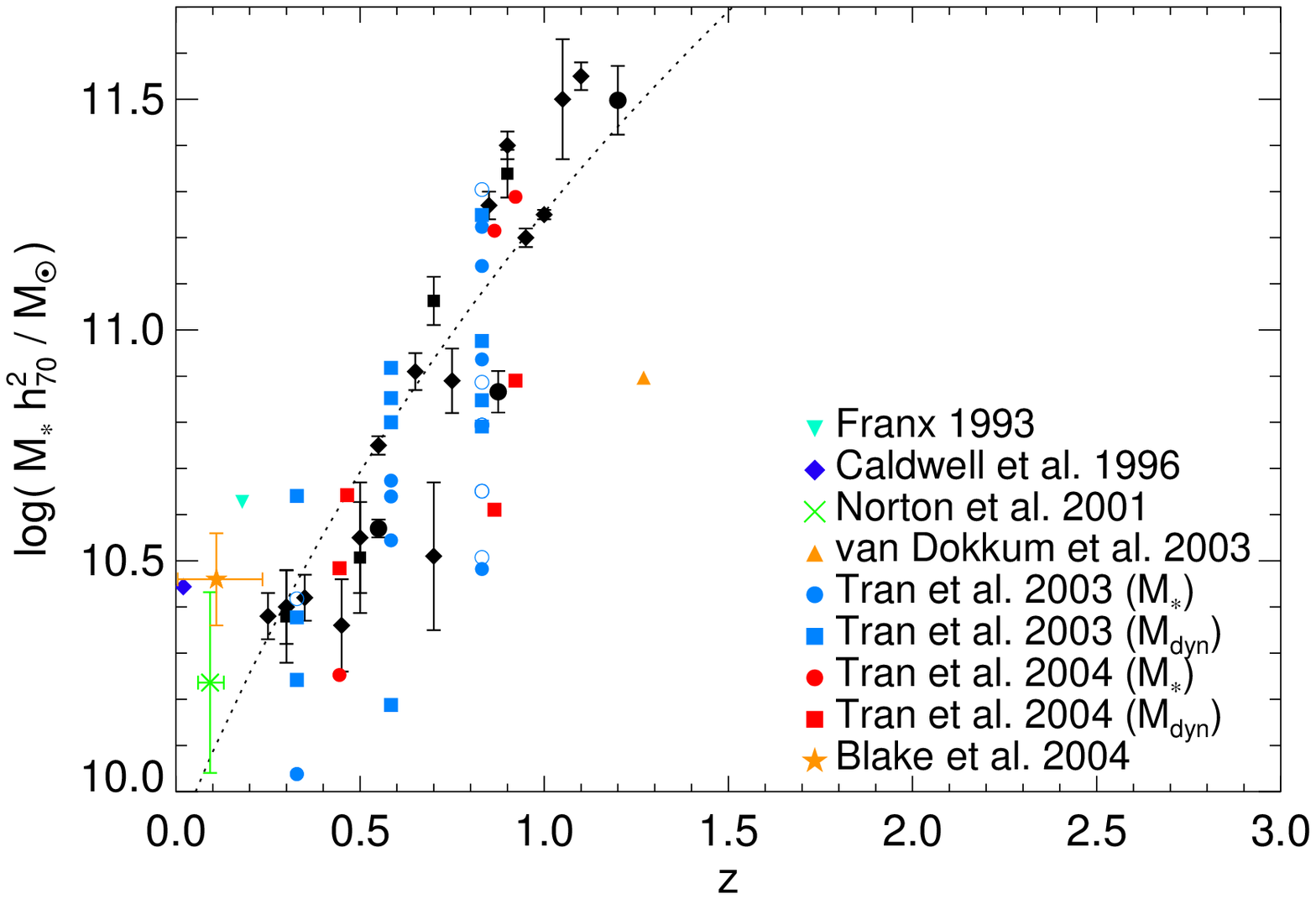}
    \caption{Observed mass of E+A galaxies (colored 
    points, as labeled) compared to the transition mass $M_{50}$
    (black points, as in Figure~\ref{fig:compare.mergers}, from color-selection) 
    as a function of redshift. We show the 
    fitted $M_{\ast}$ of the E+A mass function where 
    available (points from \citet{Norton01,Blake04} with 
    error bars, as labeled), but owing to limited samples 
    otherwise show the masses of individual E+A galaxies 
    observed in clusters \citep[][both stellar and dynamical 
    masses from \citet{Tran03} are shown, as labeled]
    {Franx93,Caldwell96,vanDokkumStanford03} and 
    in the field \citep{Tran04}. The masses of E+A galaxies appear to 
    trace the characteristic masses of mergers and the transition 
    mass as a function of redshift, but lacking larger samples at $z>0$ from which 
    to determine a full E+A mass function, it cannot be determined whether or not 
    this is merely a selection effect. 
     \label{fig:compare.ea}}
    %\epsscale{1.0}
\end{figure}
%\clearpage

Figure~\ref{fig:compare.ea} compares $M_{50}$ and the masses of observed 
E+A galaxies as a function of redshift. At low redshift, sizable samples exist, and 
we show the characteristic Schechter function $M_{\ast}$ of E+A populations. At 
higher redshift, samples are extremely limited, and we can only plot the masses 
of individual systems. The points as plotted appear to ``downsize,'' as noted in 
\citet{Tran03}, but this trend could well be completely driven by survey flux limits. 
Lacking volume-limited samples or complete E+A MFs at high redshift, we can 
only presently say that the E+A data are not {\em inconsistent} with 
the downsizing in the ``transition'' mass or any of the other hypotheses 
considered herein.

\section{The Transition Mass and Quasars}
\label{sec:quasars}

\subsection{The QLF Break Expected from the Transition Mass}
\label{sec:qlf.break.pred}

If the formation/movement of galaxies on the RS is associated with a quasar ``trigger'' 
(for example, through quasar feedback being an agent of reddening, or both being 
associated with a merger), then the observed quasar luminosity function (QLF) should 
reflect the rate of elliptical formation/buildup shown in Figure~\ref{fig:ell.buildup.mfs} -- 
indeed, in such a model, each quasar ``broadcasts'' a galaxy moving to/forming on 
the red sequence. 
(If the quasar ``lifetime'' were of order the Hubble time, of course, then the QLF would 
reflect the integrated/established early-type population, but observations constrain 
it to be much less at all redshifts, e.g.\ Martini \ 2004; Hopkins, Narayan, \&\ Hernquist 2006;  
such that the QLF tracks the rate of ``triggering.'') 

It is straightforward to compare these. A spheroid of mass $M_{\ast}$ hosts a 
black hole of mass $M_{\rm BH}=\mu\,M_{\ast}$ \citep[$\mu\approx0.001$;][]{Magorrian98,MH03}, 
confirmed by direct observations at all redshifts of interest \citep[$z\lesssim2$;][]{Shields03,Peng06,AS05a}. A ``quasar'' event is essentially defined by 
``ignition'' of the black hole for a brief time \citep[$t_{Q}\lesssim10^{7}\,$yr 
from various observations; see][and references therein]{Martini04} near 
the Eddington limit 
$L=3.3\lambda\times10^{4}\,L_{\sun}\,(M_{\rm BH}/M_{\sun})$ \citep[where
$\lambda=L/L_{\rm Edd}\approx 1$;  e.g.,][]{MD04,Kollmeier05}. Thus, the formation 
or movement to the RS of a spheroid of mass $M_{\ast}$ would be associated, 
in this scenario, with a short-lived quasar of luminosity 
\begin{equation}
\frac{L_{\rm bol}}{L_{\sun}}=33\,{\Bigl(}\frac{\lambda\,\mu}{0.001}{\Bigr)}\,
\frac{M_{\rm host}}{M_{\sun}}. 
\label{eqn:ledd}
\end{equation}
This simple effective conversion for bright quasars is supported by 
both numerical simulations of quasars and galaxy mergers \citep{H06b} 
and direct comparison of quasar and host galaxy luminosities 
\citep{VandenBerk06,Peng06,Richards06b}. 
There is, of course, some uncertainty and
observed scatter in the host galaxy-BH mass correlation and bright
quasar Eddington ratios, but it is constrained to a factor $\sim2$,
comparable to the uncertainty in the observed $\mtrans$.

If this is the dominant mode of quasar triggering, then although the 
exact normalization of the QLF (number of observed quasars) 
will depend on the ``duty cycle'' 
$\delta$ and quasar lifetime $t_{Q}$, the break $M_{\ast}$ in the host/source 
mass function (break in the rate of ``formation'' of early-type galaxies 
calculated in Figure~\ref{fig:ell.buildup.mfs}) will translate directly to 
a break $L_{\ast}$ (from Equation~\ref{eqn:ledd}) in the QLF. If $\delta$ and/or 
$t_{Q}$ are complicated functions of mass, luminosity, or redshift, they might 
change the slopes of the resulting QLF, but will not move the 
break location $L_{\ast}$. Thus, $L_{\ast}$ directly tracks the characteristic 
mass of the host population. Put another way: essentially all 
observed $L\gtrsim L_{\ast}$ quasars have $\lambda\sim1$, i.e.\ define a 
characteristic active black hole mass $M_{\rm BH}\propto L_{\ast}$, 
and since $M_{\ast}\sim M_{\rm BH}/\mu$ at all redshifts of interest, 
the characteristic host mass $M_{\ast}$ of quasars is well-defined at each $z$.

\begin{figure}
    \centering
    \figexpand
    \plotone{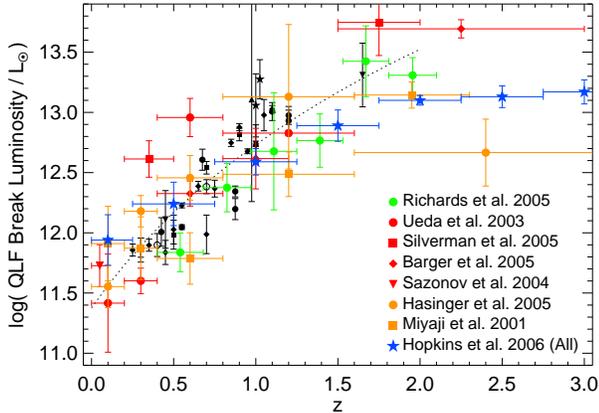}
    \caption{Predicted location of the quasar luminosity function (QLF) break $L_{\ast}$ 
    as a function of redshift, 
    from the observed transition mass (black points
    show $M_{50}$ as in Figure~\ref{fig:compare.methods}, from all samples; 
    dotted line shows best-fit trend), compared
    to the observed QLF break (colored points) from various studies. The mass $M_{50}$ is 
    converted to a luminosity assuming that the characteristic luminosity $L_{\ast}$ of quasars 
    traces their host masses, since these objects are all observed to be near-Eddington 
    and the black hole-host mass relation evolves weakly with redshift to $z\lesssim1$ 
    (alternatively, convolving the mass distribution being ``added'' to early-type populations 
    in Figure~\ref{fig:ell.buildup.mfs} with some probability of seeing the black hole in each system 
    at a given luminosity extending to $\sim L_{\rm Edd}$ yields a similar $L_{\ast}$).
    The QLF measurements shown from hard X-ray (red), soft X-ray (orange), and optical (green) 
    are converted to bolometric luminosities with the observationally determined 
    bolometric and dust corrections in \citet{HRH06}. The 
    bolometric break luminosity directly fitted in \citet{HRH06} 
    from the compilation of the samples shown and $\sim30$ other measured quasar 
    luminosity functions (see references therein) is also shown (blue). 
    The observed break luminosity from all samples is consistent with 
    the expectation of any model in which 
    the objects in Figure~\ref{fig:ell.buildup.mfs} (objects being added or 
    moving to the early-type population) are associated with quasar ``triggers,'' 
    with the black holes appropriate for their stellar mass $M_{50}(z)$ 
    briefly accreting near Eddington. 
    \label{fig:qlf.break}}
    %\epsscale{1.0}
\end{figure}

Figure~\ref{fig:qlf.break} compares the QLF characteristic luminosity 
expected (Equation~\ref{eqn:ledd}) 
from objects of the ``transition'' mass ($M_{50}$) and the observed 
characteristic QLF ``break'' luminosity $L_{\ast}$ as a function of redshift. 
The break $L_{\ast}$ is determined in the standard fashion, fitting 
the observed QLF at each redshift to a double power-law. We show $L_{\ast}$ 
measured from optical, soft X-ray, and hard X-ray studies, each converted 
(to enable direct comparison) to a bolometric luminosity $L$ using a 
standard observationally-derived bolometric correction (template quasar SED) 
and reddening correction \citep[][and references therein; note that 
adopting the less recent bolometric corrections from Elvis et al.\ 1994 or 
Marconi et al.\ 2004 yields nearly identical results]{HRH06}. We also 
show the break determined by \citet[][see their Table~2]{HRH06}, who compile a large 
number of QLF measurements through the mid and near-IR, optical, near-UV, 
soft and hard X-ray, and soft gamma ray and use these to directly 
determine the bolometric QLF.
In any case, the observed $L_{\ast}$ is robust; in fact, the (typical factor $\sim2$) 
discrepancies in $L_{\ast}$ owe mostly to the data binning and fitting function, and a direct 
comparison of the data in \citet{HRH06} shows they trace a similar turnover/break. 
We also note that the existence of a break is unambiguous 
(detected at $\gg10\sigma$ in most of the samples in Figure~\ref{fig:qlf.break}), 
regardless of whether it is sharp (as expected for a double power-law fit) or 
exhibits some higher-order curvature 
\citep[as for a Schechter function; e.g.,][]{Wolf03,Richards05}. 
It is unaffected by questions of completeness, 
as the X-ray surveys are typically 
complete to $\sim2$ orders of magnitude in luminosity below $L_{\ast}$ \citep[e.g.,][]{HMS05}.
The break luminosity also 
increases with redshift (at least to $z\sim2$), as has long been recognized in quasar surveys 
\citep[recently, e.g.][significant in each case 
at $\gg6\sigma$]{Boyle00,Ueda03,Croom04,Richards05,HMS05}, 
regardless of higher-order subtleties implied by ``luminosity-dependent density 
evolution'' \citep[e.g.,][and references therein]{HMS05} models and 
other changes in the detailed QLF shape as a function of redshift 
\citep[e.g.,][]{Richards06,HRH06}.

%\clearpage
%\clearpage

That the expected $L_{\ast}$ from $M_{50}$ agrees with the observations 
is not surprising. \citet{H06d} combined observed merger mass functions with 
a large suite of corresponding hydrodynamical merger simulations 
\citep{Robertson05b} including black hole accretion and feedback, 
to calculate what the resulting merger-driven 
QLF should be in each band at each corresponding redshift. Their predicted 
merger-driven QLF agrees well ($\chi^{2}/\nu\lesssim1$) with that observed 
at every redshift at which this comparison is possible, and the inverse (predicted 
merger MF from the QLF) also agrees well with the observations. Since we have 
shown that the buildup of the elliptical MF and $M_{50}$ trace a 
similar mass distribution to mergers, it is expected that the resulting 
$L_{\ast}$ will agree with the QLF. However, our comparison here, unlike \citet{H06d}, 
is model-independent, based only on the well-determined black hole-host mass 
relation and Eddington limit.

\subsection{The Transition Mass from the QLF Break}
\label{sec:qlf.to.mtrans}

\begin{figure*}
    \centering
    %\epsscale{1.0}
    \plotone{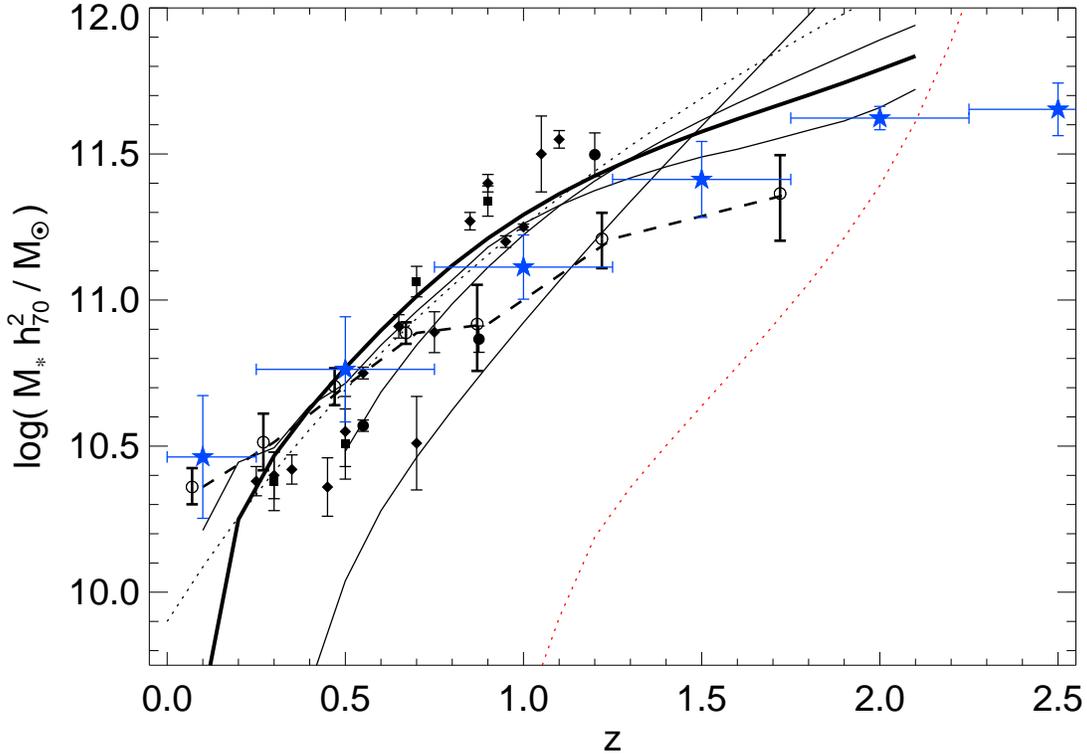}
    \caption{Observed ``transition mass'' (black points show 
    $M_{50}$ from Figure~\ref{fig:compare.methods}, from color-selected 
    samples only, for clarity; dotted black line shows best-fit trend) 
    compared with that predicted from quasar luminosity functions. 
    Blue stars show the observed QLF characteristic luminosity 
    $L_{\ast}$ from \citet{HRH06} (see Figure~\ref{fig:qlf.break}, 
    other samples not shown for clarity but trace a similar trend) directly converted to 
    a characteristic ``associated'' mass given the observed Eddington ratios 
    and black hole-host mass relation at these redshifts. 
    Solid lines show the prediction of the simple assumption that the {\em initial} trigger 
    of each bright quasar is associated with the formation/addition of an 
    early type galaxy. In detail, the compilation of quasar luminosity function 
    data from $z=0-6$ in \citet[][and references therein]{HRH06} is converted therein
    to a mass-dependent ``rate'' of early-type formation/additon, given this assumption and 
    the model Eddington ratio distributions (light curves) from Hopkins et al.\ (2005a-e,\,2006a-d). 
    Thin lines use different model fits of this ``rate''' 
    to the QLF data (and give an approximate idea of 
    the uncertainties in this empirical modeling) to calculate $M_{50}$, 
    with the thick line adopting the best-fit to the quasar data
    \citep[see][Table~5]{HRH06}. Open circles and dashed line calculate $\mtrans$ 
    (adopting the observed late-type galaxy mass functions from \citet[][z=0.05]{Bell03}, 
    \citet[][z=0.2-1.0]{Borch06}, and \citet[][z=1.0-1.7]{Fontana04})
    instead of $M_{50}$, with the same method. Red dotted line calculates $M_{50}$ 
    in this manner, but instead adopts an unphysical ``light bulb'' quasar light 
    curve model (alternatively, this assumes that all observed low-luminosity quasars 
    are in {\em ongoing} mergers/quenching). As in Figure~\ref{fig:qlf.break}, 
    the QLF break appears to trace the same parent population mass 
    and evolution with redshift as $M_{tr}$/$M_{50}$, consistent with 
    quasar {\em triggering} and the buildup of black hole mass being associated 
    with the buildup of spheroid populations. 
    \label{fig:compare.quasars}}
    %\epsscale{1.0}
\end{figure*}

In Figure~\ref{fig:compare.quasars} we invert this comparison, 
and estimate the characteristic ``transition'' mass expected 
based on the observed QLF. We first show points as in Figure~\ref{fig:qlf.break}, 
estimating a characteristic host mass from the QLF $L_{\ast}$ 
(inverting Equation~\ref{eqn:ledd}). 
However, a proper calculation is 
not so trivial, as in detail $M_{50}$ will be determined by the integrated buildup 
of the early-type MF (and late-type MF for $\mtrans$, $M_{Q}$), and 
therefore requires that we adopt some model for quasar light curves and triggering. 
Hopkins et al.\ (2006a,b) use a large set of several hundred 
hydrodynamical simulations \citep{Robertson05a,Robertson05b} of galaxy mergers, 
varying the relevant physics, galaxy properties, orbits, and system masses, 
to quantify the quasar lifetime (and related statistics) as a
function of the quasar luminosity. They define the quantity $t_{Q}(L\,|\,M_{\rm BH})$, 
i.e.\ the time a quasar of a given BH mass (equivalently, peak quasar 
luminosity $t_{Q}(L\,|\,L_{\rm peak})$) will be observed at a given luminosity $L$. They 
further demonstrate that this quantity is robust across the wide range of varied 
physics and merger properties; for example, to the extent that the final 
BH mass is the same, any ``major'' 
merger of sufficient mass ratio (less than $\sim5:1$) will produce an identical effect. 
Given the tight black hole-host mass relation ($M_{\rm BH}-M_{\ast}$), it is 
trivial to write this as $t_{Q}(L\,|\,M_{\ast})$. Since at all $L$, $t_{Q}\ll t_{H}$ (the Hubble time), 
the observed QLF $\phi_{Q}(L)$ is given by
\begin{equation}
\phi_{Q}(L,\,z) = \int t_{Q}(L\,|\,M_{\ast})\,\frac{{\rm d}\Phi(M_{\ast},\,z)}{{\rm d}t\,{\rm d}\log{M_{\ast}}}\,{\rm d}\log{M_{\ast}}, 
%\phi_{Q}(L,\,z) = \int t_{Q}(L\,|\,M_{\ast})\,{\dot{\phi}(M_{\ast},\,z)}\,{\rm d}\log{M_{\ast}}, 
\label{eqn:qlf}
\end{equation}
where ${{\rm d}\Phi(M_{\ast})}/{{\rm d}t\,{\rm d}\log{M_{\ast}}}$ (hereafter $\equiv\dot{\phi}(M_{\ast},\,z)$) is the rate of 
quasar ``triggering'' as a function of host spheroid mass at a given redshift. 
If the ``trigger'' is associated with formation of the spheroid or movement of a 
galaxy to the RS, then $\dot{\phi}(M_{\ast},\,z)$ {\em is} 
the rate of buildup of the RS as a function of mass, directly comparable to 
that in Figure~\ref{fig:ell.buildup.mfs}. 

%\clearpage
%\clearpage

Knowing $\phi_{Q}(L)$ directly from observations and adopting the 
\citet{H06a,H06b} $t_{Q}(L\,|\,M_{\ast})$ (which is at least consistent with all 
quasar observational constraints), the inversion of Equation~(\ref{eqn:qlf})
yields $\dot{\phi}(M_{\ast},\,z)$. 
\citet{HRH06} perform this inversion, using their large compilation of observed $\phi_{Q}(L)$, 
and quote the best-fit $\dot{\phi}(M_{\ast},\,z)$ (Table~5). We adopt their 
best-fit model for each redshift interval (individually; although a global fit 
yields similar results). 
If, again, this represents the buildup of the RS, then we can 
integrate from $z\rightarrow\infty$ to obtain the RS MF at all redshifts and 
calculate $M_{50}(z)$. The late-type MF is reasonably well-measured over the 
range of interest, so comparing it with this integration also yields an expected 
$\mtrans$ and $M_{Q}$. We compare 
these estimates with the observed $M_{50}$ in 
Figure~\ref{fig:compare.quasars}, and find they agree at all observed redshifts. 

\begin{figure*}
    \centering
    \figexpand
    \plotone{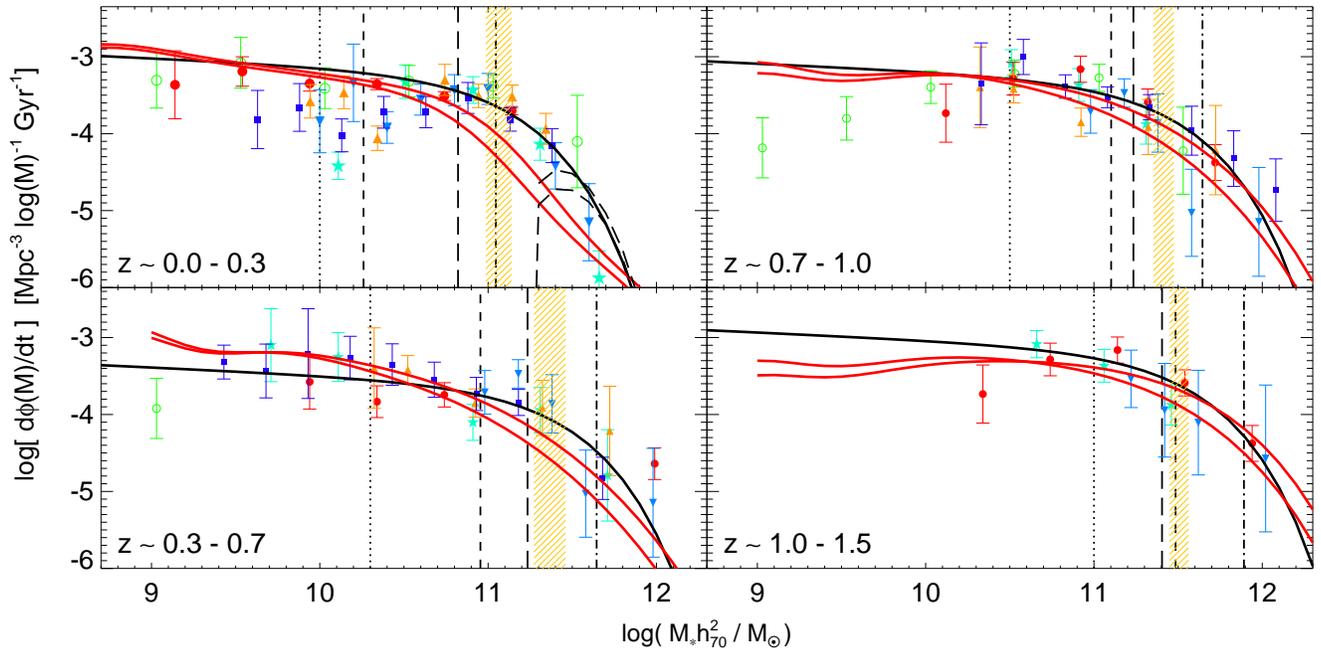}
    \caption{The time-averaged rate of buildup of early-type populations, as in 
    Figure~\ref{fig:ell.buildup.mfs}, compared to that implied by the 
    quasar luminosity function (QLF) if every quasar ``trigger'' is associated 
    with the movement/formation/transition to the red sequence
    of a corresponding $(M_{\ast}\sim10^{3}\,M_{\rm BH}$)
    spheroid (red lines). The functional form for the QLF and implied 
    ``triggering rate'' is taken from the best-fit given in \citet{HRH06}. Lower 
    line in each panel assumes instantaneous reddening, upper line a $1\,$Gyr delay 
    (post-quasar peak) before 
    objects appear on the red sequence. The implied  
    rate, if quasars and the blue-red transition are associated with 
    the same event, agrees well with the buildup of 
    elliptical populations at all masses at moderate and high redshifts. At low 
    redshift $z\lesssim0.3$ (and to a lesser extent, at $z\sim0.3-0.7$), 
    the implied rate from the QLF falls below the observed buildup 
    at high masses. The deficit can be accounted for at the highest masses with 
    the observed rate of ``dry'' mergers (which, by definition, will not generally 
    trigger quasar activity).
    \label{fig:ell.buildup.vs.qso}}
    %\epsscale{1.0}
\end{figure*}

Having obtained the rate of ``buildup'' of early type MFs expected if 
each quasar ``trigger'' is associated with the formation/movement of a 
RS galaxy, we can directly compare with the rate of buildup implied by 
observed early-type MFs. Figure~\ref{fig:ell.buildup.vs.qso} plots the 
time-averaged buildup determined from the QLF, from the same 
$\dot{\phi}(M_{\ast},\,z)$ as Figure~\ref{fig:compare.quasars}, 
compared with the observed buildup from Figure~\ref{fig:ell.buildup.mfs}. 
We consider both the mean time-averaged buildup (averaged over each appropriate 
redshift interval) assuming each quasar ``trigger'' 
is instantaneously associated with the movement of a galaxy to the RS, and 
that expected if there is a uniform $\sim1$\,Gyr delay after each quasar before 
the galaxy becomes red (allowing time for e.g.\ gas exhaustion and reddening). 
In either case, this estimate agrees with the observed 
``buildup'' of elliptical populations, at all masses and redshifts ($\chi^{2}/\nu\sim1$ 
at all $z\gtrsim0.3$). The latter ($1$\,Gyr delay) 
case gives marginally better agreement, but the difference between the two 
is comparable to the uncertainties in either determination of 
$\dot{\phi}(M_{\ast},\,z)$ (see Figure~\ref{fig:compare.quasars}). 
At the highest masses at low redshifts ($z\lesssim0.3$; although also 
to a lesser extent at $z\sim0.3-0.7$), this estimate falls short of observed rates of 
``buildup.'' However, this is precisely where we have estimated that observed rates of 
``dry'' mergers can account for early-type growth. Since gas-free mergers are not expected to 
trigger quasar activity, it is not surprising that this would not be implicit in quasar 
luminosity functions. Allowing for the contribution of dry mergers shown 
at $z\lesssim0.3$ improves the agreement considerably ($\chi^{2}/\nu\sim2$). There is 
still some tension matching the observations near $10^{11}\,M_{\sun}$, 
but it is important to note that at these redshifts, the cosmic variance associated with 
small volume, narrow-field galaxy surveys and, perhaps more importantly, with even 
wide-field quasar surveys (given the very low local space density of 
quasars), is largest. 

%\clearpage
%\clearpage

Having estimated the rate of quasar ``triggers,'' 
$\dot{\phi}(M_{\ast},\,z)$, in Figure~\ref{fig:ell.buildup.vs.qso}, 
then if each such ``trigger'' is 
in fact a galaxy merger, we can convert this to an expected merger 
fraction in exactly the same manner as we converted the rate of early-type 
``buildup'' in Figure~\ref{fig:ell.buildup.mfs} (i.e.\ simply assuming an 
observable merger timescale $t_{\rm merger}$). This is shown in Figure~\ref{fig:merger.frac}, 
along with the observed additional contribution from dry mergers. Given the 
agreement with the rate of elliptical ``buildup'' in Figure~\ref{fig:ell.buildup.vs.qso}, 
it is not surprising to find this agrees with observed merger fractions.

As a caution, we should note that these calculations can give a misleading 
result if the full luminosity dependence of the quasar ``lifetime'' 
from simulations \citep{H05a,H05b,H06b} and observations \citep[e.g.,][]{AS05b,Volonteri06} 
is not properly taken into account. Such a case is not, of course, 
well-motivated physically, although it may represent alternative 
quasar feedback models (or a complete lack of such feedback), 
but it is nevertheless sometimes adopted for simplicity.
Why should such a simplified model give a 
qualitatively different result? In the $t_{Q}(L\,|\,M_{\rm BH})$ model 
we consider, $t_{Q}$ is larger at low luminosities, because low-level AGN 
activity can persist for a long time after the violent, sudden high-accretion rate 
episode in a merger. Ignoring this luminosity dependence and assuming, e.g.\ 
that all quasars turn on and off (as ``light-bulbs'') for a short time implies that 
all observed quasars, even those at very low luminosity, are seen at 
(or very near) their ``trigger,'' i.e.\ are in {\em ongoing} mergers. This gives a 
misleading estimate of the number of mergers needed to account for 
the QLF, and as a result yields an incorrect estimate of host 
luminosity functions and black hole mass functions \citep[e.g.,][]{H06b,H06d}, 
as well as, consequently, erroneous estimates of the associated 
``transition'' mass.

\subsection{Further Tests of This Association}
\label{sec:tests}

\begin{figure*}
    \centering
    %\epsscale{1.1}
    \plotone{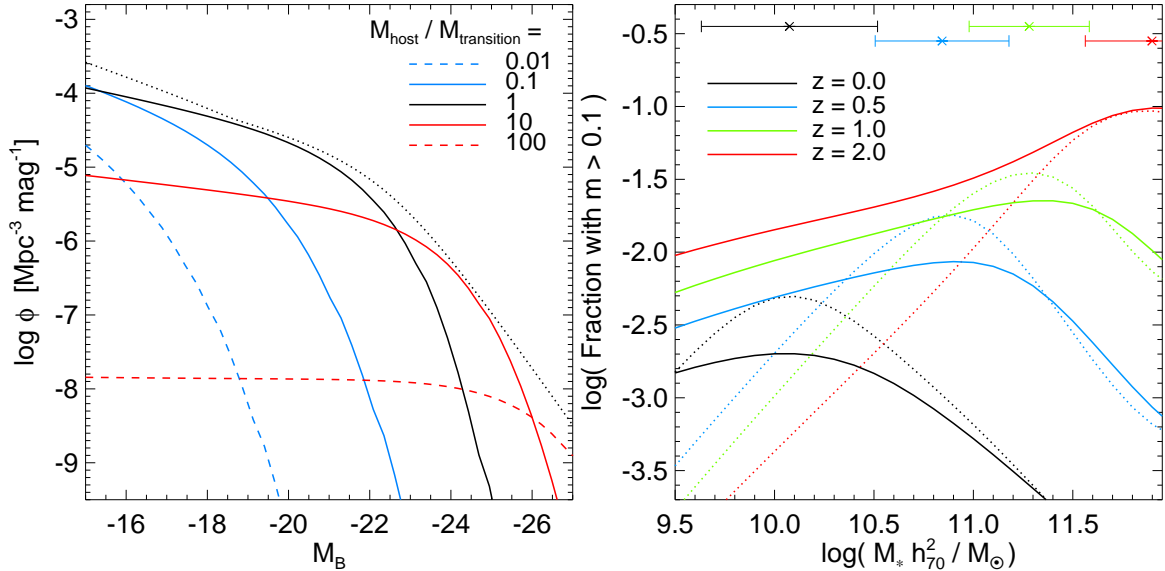}
    \caption{{\em Left:} Predicted contribution to the total $B$-band QLF (dotted, 
    from the compilation of \citet{HRH06})
    from host galaxies in different 
    mass intervals relative to the observed transition mass as labeled, from 
    the models of Hopkins et al.\ (2005a-e,\,2006a-d; see Table~5 of 
    \citet{HRH06}) (shown at $z=0.5$, but qualitatively 
    similar at all redshifts of interest). 
    {\em Right:} Predicted fraction of galaxies hosting 
    an AGN with an accretion rate (relative to Eddington) $\dot{m}>0.1$, as a function 
    of mass, at different redshifts as labeled. Solid lines show the best-fit 
    model from \citet[][Table~5]{HRH06}, dotted lines the $2\sigma$ minimum 
    allowed contribution to the QLF from low-$M_{\ast}$ hosts. 
    Points of the corresponding color show the 
    observed ``transition'' mass at each redshift (from the best-fit trend; dashed 
    line in Figure~\ref{fig:compare.methods}; error bars show approximate 
    dispersion from different ``transition'' mass definitions). 
    Although systematic uncertainties at low masses are 
    large, the predictions above -- that quasar $\sim L_{\ast}$ hosts 
    and a turnover or peak in the ``active'' galaxy fraction should correspond 
    to $\sim\mtrans$ hosts -- are robust expectations of models which associate a 
    blue-red transition and quasar activity. 
    \label{fig:quasar.tests}}
    %\epsscale{1.0}
\end{figure*}

Having considered the \citet{H06b} models of merger-triggered quasar lightcurves, 
we briefly note additional future tests of these models and the generic association 
between the blue-red transition or elliptical formation and quasar activity. 
In Figure~\ref{fig:quasar.tests} we compare the observed QLF 
with the expected ``conditional'' QLF, i.e.\ the contribution to the QLF 
from hosts/merger remnants with different masses relative to the 
observed ``transition'' mass. In other words, the contribution 
\begin{equation}
\Delta\phi(L) = t_{Q}(L\,|\,M_{\ast})\,\dot{\phi}(M_{\ast},\,z)\,
{\Delta}\log{M_{\ast}}
\end{equation}
from Equation~(\ref{eqn:qlf}). 
The QLF near $L_{\ast}$ corresponds to objects with $M_{\ast}\sim M_{50}$. 
At the faintest and brightest luminosities, there are contributions from 
smaller and larger hosts, respectively 
(and a significant fraction of objects at the lowest luminosities will not 
necessarily be merger-triggered; \citet{HH06}),
but it is clear from the figure that a direct measurement of the host
masses of characteristic quasars at $z$ should find their hosts
dominated by objects with $M_{\rm host}\sim \mtrans$ or $M_{50}$, 
many of which should appear as relatively young ellipticals, if this picture 
is correct. 

%\clearpage
%\clearpage

In this scenario, one might also expect that objects around the 
``transition'' mass preferentially host AGN. Since Equation~(\ref{eqn:qlf})
implicitly defines the probability to see a galaxy with mass $M_{\ast}$ 
(BH mass $M_{\rm BH}\sim\mu M_{\ast}$) at 
luminosity $L$, we can estimate the fraction of such galaxies with a 
given Eddington ratio. Figure~\ref{fig:quasar.tests} plots 
the fraction of galaxies hosting an AGN
with an accretion rate (relative to Eddington) $\dot{m}>0.1$ as a
function of galaxy stellar mass at several redshifts, compared to the 
observed ``transition'' mass at that $z$. (Note the actual 
AGN luminosity will be $\dot{m}\,L_{\rm Edd}(M_{\rm BH})$.) 
This particular prediction is sensitive to the different 
fits to the ``triggering'' rate 
$\dot{\phi}(M_{\ast},\,z)$ 
provided in \citet{HRH06}, especially at low mass ($M\ll M_{50}$), but 
the trend that the peak/turnover in this distribution tracks the 
``transition'' mass is robust.

\section{The Transition Mass and the Halo Quenching Mass}
\label{sec:halomass}

In most semi-analytic models, gas infalling in dark matter 
halos is shock-heated to the virial temperature, and, in low mass 
halos, subsequently cools on a short timescale, allowing rapid accretion 
onto the central halo galaxy and defining a ``rapid cooling'' or 
``cold accretion'' regime. 
However, in massive halos, the 
cooling time is longer and gas forms a quasi-static ``hot'' halo, defining a 
``static hot halo'' or ``hot accretion'' regime 
\citep[e.g.,][]{ReesOstriker77,Blumenthal84}. More recently, it has been 
suggested that the transition between these regimes is sharp, near a 
halo mass $\mhaloquench\sim10^{12}\,M_{\sun}$ (although this 
number is uncertain by a factor of several) at low redshift
\citep{Birnboim03,Keres05}, 
and that suppression of future cooling and accretion is 
very efficient, essentially ``cutting off'' all gas supplies above this mass 
\citep[e.g.,][]{Dekel06} and ``quenching'' star formation. 

The shock-heating of infalling gas need not be the {\em specific} physical agent of 
this ``quenching'': for example ``radio mode'' or low-luminosity, continuous AGN 
feedback \citep{Croton05,Cattaneo06} or cyclic, short-lived quasar activity \citep{Binney04} 
may be invoked to maintain the gas in the ``hot'' phase. There is therefore a potentially 
important distinction between semi-analytic models (SAMs) which 
assume that the feedback mechanism is ``at ready,'' such that 
upon crossing the critical mass $\mhaloquench$, star formation and gas accretion 
onto the central galaxy is {\em instantaneously} terminated, and those that 
require some additional mechanism or process (such as the formation of a 
relatively massive bulge and black hole) to drive the blue-red transition and 
transformation/movement of galaxies to the RS. 

This essentially relates to the important distinction, discussed in \S~\ref{sec:intro}, between the 
mechanism by which galaxies {\em become} red/elliptical and that 
by which they {\em maintain} their colors/low star formation rates. The 
key value of invoking this ``hot accretion'' regime in SAMs has been 
the ability to suppress star formation on timescales of order the Hubble time. However, 
although this could, in principle, be {\em necessary} to yield red galaxies at $z=0$, 
it does not automatically follow that it is {\em sufficient}. In other words, 
there may be other processes (e.g.\ mergers and/or quasars) which drive the 
blue-red transition and movement to the RS, and the 
``hot accretion'' mode simply maintains these galaxies at their low star formation rates. 

One possible interpretation of the observed ``transition mass,'' perhaps the most naive, 
is that the transition mass simply 
represents the stellar mass hosted in $\mhaloquench$ halos at each redshift. 
If we adopt the expected halo quenching mass $\mhaloquench$ from 
\citet{Dekel06}, and either assume the galaxies hosted have the same stellar mass 
as those in $z=0$ halos of the same mass \citep[measured in][]{Mandelbaum05}, 
or that they are already 
fully assembled (i.e.\ have stellar masses at $z$ appropriate for what their 
halo mass will be at $z=0$), we can compare with our observed ``transition mass.'' 
We find that while the two are similar at low redshifts, they diverge at higher-$z$. 
This is, of course, where the observations are most uncertain, so it may simple reflect 
a systematic error in our estimation of the ``transition mass.'' But it probably also 
reflects the possibility that, in these models, the ``transition mass'' has a more complex 
physical origin than simply tracing $\mhaloquench$. As noted in \S~\ref{sec:compare.mstar}, 
allowing for more complex and realistic distributions of galaxies in transition to the RS 
can affect quantities such as the ``break'' mass $M_{\ast}$ and ``transition'' mass 
in a non-trivial manner. But there is also the possibility that the ``transition'' to the RS 
requires additional processes beyond the initial cutoff of new gas supplies in 
``hot mode'' accretion, such as gas exhaustion, mergers, and/or quasar 
activity to operate, which are what we see traced by the observed transition mass. 

%\clearpage
\begin{figure*}
    \centering
    \figexpand
    \plotone{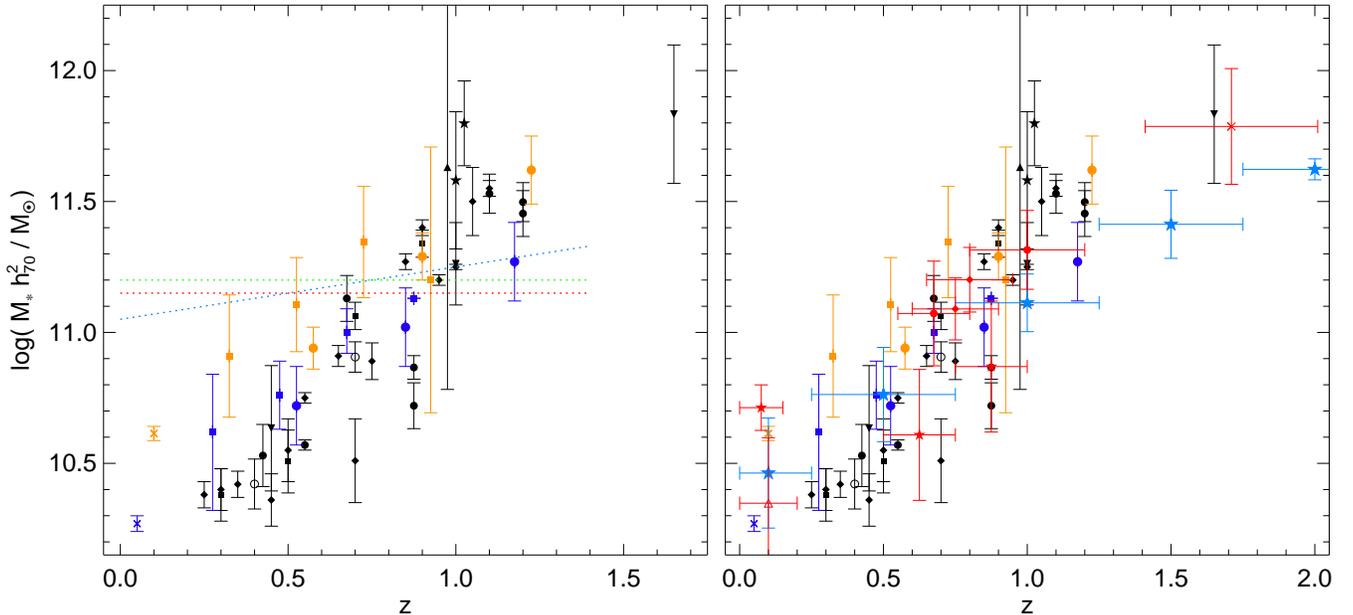}
    \caption{{\em Left:} The ``transition'' mass determined by various 
    definitions ($M_{50}$ from all samples, $\mtrans$ from color-selected samples, 
    and $M_{Q}$ from color-selected samples shown as black, purple, and orange points, 
    respectively, as in Figure~\ref{fig:compare.methods}), as a function of redshift. 
    Dotted lines show the characteristic mass $M_{\ast}$ of all, red, and blue galaxies 
    (green, red, and blue, respectively; see Figure~\ref{fig:compare.mcrit.mstar}). 
    The null hypothesis that ``transition'' mass objects (as well as quasars and mergers; 
    see right panel) are drawn randomly/uniformly from the all, red, or blue galaxy 
    population can be ruled out at $>6\sigma$ ($>5\sigma$ for quasars, $>3\sigma$ 
    for mergers).     
    {\em Right:} The ``transition'' mass, as in the left panels, together with the 
    characteristic masses $M_{\ast}$ of merger mass functions (red points; as 
    in right panel of Figure~\ref{fig:compare.mergers}) and quasar 
    hosts (blue stars; as in Figure~\ref{fig:compare.quasars}). The observations at 
    all redshift are consistent with the hypothesis that mergers, quasars, and 
    the transition/addition to the red sequence are associated with the 
    same event. 
    \label{fig:compare.models}}
    %\epsscale{1.0}
\end{figure*}
%\clearpage

It is also worth considering whether or not the flow of galaxy host halos across $\mhaloquench$ 
is consistent with the number and mass densities of ``transition'' objects and the 
buildup in early-type populations which we have estimted 
from the observations. 
The rate at which halos cross a given mass threshold $M$ is 
straightforward to calculate in linear theory from the Press-Schechter formalism, 
\begin{equation}
F(>M\,|\,z) = {\rm erfc}\,{\Bigl(}\frac{\delta_{\rm coll}(z)}{\sqrt{2}\sigma(M)}{\Bigr)}, 
\end{equation}
and adopting either a simple mean stellar-to-dark matter mass ratio 
\citep[from the calibration of][]{Mandelbaum05}, or integrating (at this halo mass) over 
the population of inferred hosted galaxies from the conditional galaxy 
mass functions (CMFs; i.e.\ probability that halos of mass $M$ host galaxies of stellar mass 
$M_{\ast}$) yields an estimate of the rate at which stellar mass crosses this threshold. The local 
CMF is determined (albeit indirectly) entirely from observations 
of galaxy mass/luminosity functions and clustering 
\citep[e.g.,][]{Yang03,Yang05,Zheng05}, and has been subsequently 
measured directly in 2dFGRS group catalogues by \citet{Yang05}, 
and is well-constrained with typical uncertainties smaller than or comparable 
to those in our estimate of the rate of early-type ``buildup'' 
(at least for $M_{\ast}\gtrsim10^{10}\,M_{\sun}$ of interest here; the 
MF at lower masses depends on the mass threshold for 
inclusion of satellite systems). \citet{Yan03,Cooray05,Cooray06} 
extend the CLF/CMF to high redshifts ($z\lesssim4$) 
using a large number of luminosity function and clustering estimates 
from wide-area surveys.
Note that \citet{Yang03,Yang05} and \citet{Cooray06} actually measure the 
conditional luminosity function (CLF); we convert to a CMF using the appropriate $M/L$ ratios 
as a function of mass from \citet{Bell03}, and assume these $M/L$ values evolve with 
redshift following the best-fit stellar population models as a function of mass from 
\citet{Gallazzi06} and \citet{Renzini06}. Checking directly \citep[following 
the methodology of ][]{Yang03} shows that 
this agrees with the \citet{Bell03} mass functions (see also their Figure~19)
and the \citet{Li06} measured clustering as a function of stellar mass, and furthermore,  
these CMFs agree well with those directly determined in \citet{Zheng05}. 
Ultimately, there are a number of systematic (factor $\sim2-3$) 
uncertainties in this comparison, and 
our (admittedly crude) empirical calculation ignores the fact that, in ``quenching'' 
models, crossing the quenching threshold itself may change the stellar-to-dark matter 
mass ratios and stellar $M/L$ values. However, within these rather large uncertainties, 
our purely empirically estimated rate at which galactic host halos cross 
$\mhaloquench$ is consistent with the possibility that this is a necessary 
prerequisite for ``transition'' to the RS.

\section{Clustering: An Independent Test}
\label{sec:clustering}

We compare the populations we have considered in an independent manner 
by examining their clustering properties.
If a population (i.e.\ a given set of ``parent'' halos) clusters 
with a given bias $b(z)$ at some redshift $z$, then the subsequent evolution 
in their bias is trivially calculated in linear theory  
\begin{equation}
b(z=0) = 1 + {D(z)}\,[b(z)-1], 
\label{eqn:bias}
\end{equation}
where $D(z)$ is the growth factor 
\citep{Croom01}, regardless of the processes (accretion, mergers, etc.) that 
affect the halos (and galaxies) themselves. The bias 
of galaxies (specifically red/elliptical galaxies) 
as a function of stellar mass is well-determined at $z=0$ \citep[we adopt 
the recent determination from the SDSS in][with typical $\lesssim10\%$ uncertainty]{Li06}, so 
given the bias of a population at $z$ and evolving it to $z=0$ with 
Equation~(\ref{eqn:bias}) yields the characteristic $z=0$ stellar mass 
of this population (i.e.\ the average stellar 
mass of which the population is the ``parent''). 

\begin{figure*}
    \centering
    \figexpand
    \plotone{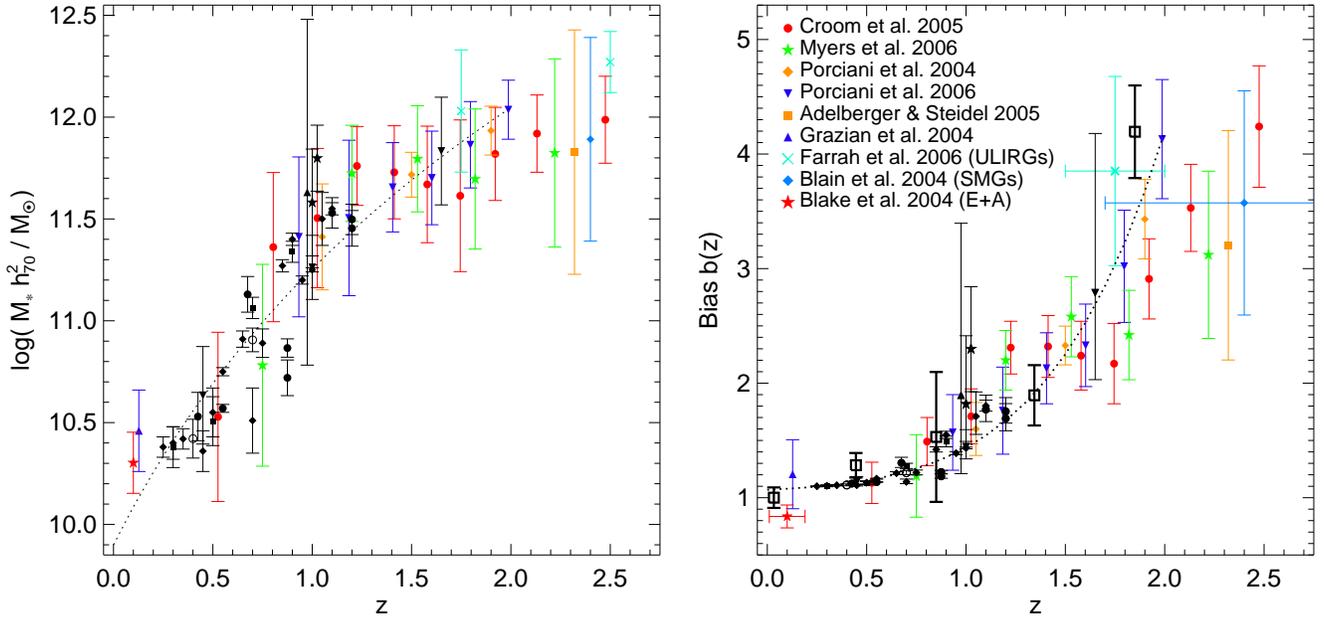}
    \caption{
    {\em Left:} Observed ``transition'' mass (black points show $M_{50}$  
    as in Figure~\ref{fig:compare.methods}, from all samples; black dotted line shows 
    best-fit trend) as a function of redshift, 
    compared with the characteristic host masses (colored points) 
    of quasars, recently-formed 
    elliptical (E+A/K+A) galaxies, and ongoing bright mergers (ULIRGs/SMGs), 
    estimated from their clustering properties. 
    {\em Right:} Corresponding bias as a function of redshift. 
    Black points as in left panel show $b(z)$ calculated from $M_{50}(z)$ 
    and the observed $z=0$ bias \citep{Li06} for that stellar mass (i.e.\ assuming passive 
    evolution), open black squares show $b(z)$ determined directly from 
    observations for red galaxies with the appropriate $M_{50}(z)$ 
    \citep[$M_{50}(z)$ from best-fit trend; points of increasing redshift from][respectively]
    {Li06,Shepherd01,Pollo06,Meneux06,Brown05}. 
    Colored points show $b(z)$ observed for the quasar, E+A, and merger populations, as labeled. 
    Note that $b(z)$ is defined from the clustering length $r_{0}$ and thus the relative 
    bias and $M_{\ast}$ shown are only weakly cosmology-dependent. 
    This provides a completely independent check 
    of the previous comparison between these populations, but one 
    which suggests a similar co-evolution.
    \label{fig:clustering}}
    %\epsscale{1.0}
\end{figure*}

Figure~\ref{fig:clustering} shows this $M_{\ast}$, calculated from various 
clustering measurements $b(z)$ of quasars, mergers (ULIRGs and SMGs), 
and E+A galaxies as a function of redshift, and compares to the 
``transition'' mass $M_{50}$ at each redshift. 
We invert this as well; 
knowing $M_{50}(z)$, evolve $b(M_{50},\,z=0)$ with 
Equation~(\ref{eqn:bias}) to estimate $b(z)$. 
Note that all $b(z)$ shown from measurements are converted from the directly observed 
clustering length $r_{0}$, which for a power law correlation function 
yields 
\begin{equation}
r_{0}=r_{0}(z=0)\,[b\,D(z)]^{\gamma/2}, 
\end{equation}
with $r_{0}(z=0)\sim5\,h^{-1}\,{\rm Mpc}$ and $\gamma\sim1.8$ \citep[see, e.g.,][]{Norberg02}. 
The absolute value of the bias as plotted is then weakly dependent on cosmology
(and this conversion, of course, is inexact), 
but the important point for our purposes is that the {\em relative} bias of all points 
plotted (and $b(M_{\ast},\,z=0)$ with which we compare) is insensitive to the cosmology.

%\clearpage
%\clearpage

These comparisons do assume that the stellar mass of individual systems 
does not change much from $z$ to $z=0$, i.e.\ that, once formed, ellipticals 
are passively evolving. However, we can easily eliminate this assumption, 
by considering the clustering directly observed for red galaxies of mass $M_{50}(z)$ 
at that redshift, and Figure~\ref{fig:clustering} shows this as well. In either 
case, the agreement with the clustering of quasars and (albeit much less 
well-constrained) merger/E+A populations is good. 
This also agrees with determinations of e.g.\ the typical overdensities 
and small-scale clustering of 
quasars and ULIRGs \citep{Farrah04,Hennawi06,Serber06}. 

This method by which we compare clustering is only weakly 
dependent on cosmology, through the growth factor $D(z)$ (independent of 
e.g.\ $\sigma_{8}$). There are some caveats, however. 
Technically, we are estimating the 
mass which has exactly the observed bias; this is some weighted 
mean mass. However, theoretical expectations from physically 
motivated quasar light curve models \citep{Lidz06} and direct observations 
of clustering as a function of luminosity \citep{AS05b,Croom05,Myers06}
suggest that quasar clustering depends only weakly on luminosity, 
reflecting a reasonably well-defined characteristic host mass. 
These comparisons will also, of course, be affected if the clustering of 
mergers is different on large scales from that of non-merging halos of the same mass (a 
so-called ``merger bias''). However, a number of investigations have found no such 
dependence \citep[e.g.,][]{LemsonKauffmann99,KH02,Percival03} and even 
where more recent investigations have seen such an effect \citep{Gao05} it has been 
restricted to small mass halos (below the ``collapse mass'', i.e.\ where 
$b=1$) at $z=0$, and therefore the assumption of 
no merger bias has generally been adopted 
in quasar clustering studies \citep{MartiniWeinberg01,HaimanHui01,AS05b,Croom05,Myers06,Lidz06}.

We can repeat this comparison using the formalism of \citet{MoWhite96} from 
linear collapse theory, which yields a characteristic halo mass from a 
given observed $b(z)$. We use the observed 
stellar mass-halo mass relations calibrated for elliptical galaxies 
from weak lensing measurements in \citet{Mandelbaum05} to convert these halo masses 
to a stellar mass $M_{tr}$. We convert between halo mass and bias with 
the method of \citet{MoWhite96} modified following 
\citet{ShethTormen01} in our adopted cosmology 
(in detail assuming $\sigma_{8}=0.8$, $n_{s}=0.98$) 
with the power spectrum computed following \citet{EH99}. The results are similar, 
but are much more sensitive to the adopted cosmology and systematics 
in the stellar mass-halo mass relation in this approach. 

In considering the clustering of $M_{50}$ objects, we have 
considered the directly measured bias of objects with mass $M_{50}(z)$ 
at redshift $z$, as well as the ``passively evolved'' clustering from 
the $z=0$ bias as a function of mass. 
We can gain further insight into 
the evolution of these populations by comparing the two. 
Knowing the observed bias of $M_{50}(z)$ objects at $z$, we can 
evolve this to $z=0$ given Equation~(\ref{eqn:bias}), and then use 
$b(M_{\ast}\,z=0)$ to obtain the 
typical stellar mass hosted by these systems {\em at $z=0$.} 
Comparing that to their stellar mass at $z$, namely $M_{50}(z)$, 
shows by how much the typical stellar mass of the population has grown. 
We could also estimate this in a more indirect fashion, using linear 
theory to estimate a host halo mass $M_{\rm halo}(z)$ given $b(z)$, then knowing 
the $z=0$ mass of a halo with mass $M_{\rm halo}(z)$ at $z$, use the 
local galaxy stellar-halo mass calibrations from \citet{Mandelbaum05} to 
obtain $M_{\ast}(z=0)$. Again, this approach is considerably more 
sensitive to the assumed cosmology, but in our adopted case yields similar results.

\begin{figure*}
    \centering
    %\epsscale{1.0}
    \plotone{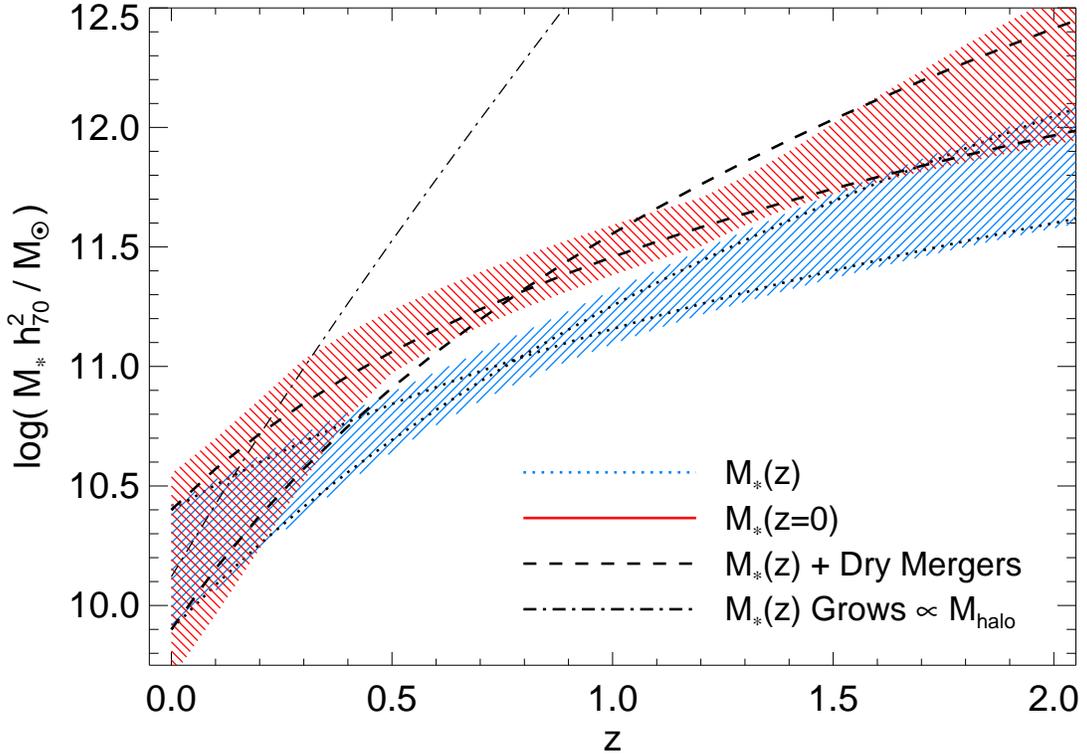}
    \caption{The best-fit ``transition'' stellar 
    mass $M_{50}(z)$ (see Figure~\ref{fig:compare.methods}; 
    dotted lines with blue shaded range show $\sim1\sigma$ range) at a given 
    redshift, compared to the inferred stellar mass of the same objects at 
    redshift $z=0$ (red shaded range). The observed bias of a fixed set of 
    objects (halos) with stellar mass $M_{50}(z)$, $b(M_{50}(z),\,z)$, 
    evolves simply to $z=0$ \citep{Croom01}, where a comparison with 
    the local $b(M_{\ast},\,z=0)$ yields the typical stellar mass hosted by such halos. 
    Dashed lines show the result of taking $M_{50}(z)$ (dotted lines) and allowing 
    for a constant rate \citep[$\sim1$ since $z=1$, as suggested by observations;][]{vanDokkum05,Bell06}
    of major (equal-mass) ``dry'' mergers. Dash-dotted line assumes that subsequent (after $z$)
    galaxy assembly tracks halo assembly (i.e.\ efficient dry merging; no ``downsizing'' in 
    {\em assembly} times), as implied by some 
    semi-analytic models \citep[e.g.,][note this is {\em not} a rigorous comparison 
    with these models]{deLucia06b,Cattaneo06}. 
    Although systematic uncertainties are sufficiently large that this should not be 
    considered evidence {\em for} a particular amount of dry merging, the 
    comparison demonstrates that the evolution in $M_{50}(z)$ is completely consistent 
    with observational evidence for a significant growth by dry mergers since $z\gtrsim1$. 
    A very large number of dry mergers is, however, inconsistent, and this also 
    violates the direct observationally determined rates of dry mergers above
    (galaxies grow by a factor $\gtrsim2$ since $z=1$). Note that dry mergers do not, 
    by definition, ``build up'' the {\em total} mass budget on the red sequence, and at 
    observed rates, have a relatively weak effect on the RS MF near the ``transition'' mass 
    (see also Figure~\ref{fig:ell.buildup.mfs}). 
     \label{fig:clustering.to.m.z0}}
    %\epsscale{1.0}
\end{figure*}

In Figure~\ref{fig:clustering.to.m.z0}, we use this to compare $M_{50}(z)$, 
the stellar mass of ``transition'' mass objects at $z$, with 
$M_{\ast}(z=0)$, i.e.\ the mean 
$z=0$ stellar mass which is typically hosted by the 
evolved ``parent halos.'' Unless ``transition'' mass objects comprise some 
unusual outlier in their halo properties, this should represent the typical 
stellar mass these objects will grow to by $z=0$. 
We compare with the expectation, following e.g.\ \citet{Bell06}, that 
these stellar masses grow at a rate corresponding to one major (mass ratio $1:1$) 
merger since $z=1$. We also consider the case if the stellar mass in these 
objects grows in fixed proportion with their host dark matter halos. 

The systematic uncertainties (and measurement errors in $b(z)$) are sufficiently large 
that we should regard these comparisons with caution, and 
{\em not} consider this as evidence for a particular ``amount'' of 
dry merging. However, the estimated $M_{\ast}(z=0)$ demonstrates that the 
$M_{50}(z)$ measurements are completely {\em consistent} with subsequent growth 
by dry mergers at observationally inferred rates. 
Growth in proportion to the host halo mass, by contrast, is extremely difficult to reconcile 
with observed properties of the galaxies. This is not surprising, as the existence of any 
significant $\sim10^{12}\,M_{\sun}$ galaxy population at $z=1$ without a corresponding 
$\sim10^{13}-10^{14}\,M_{\sun}$ galaxy population at $z=0$ implies that, at least for 
some objects assembled most rapidly, subsequent galaxy {\em assembly} must 
lag behind halo growth (or subsequent growth in these halos must be 
anomalously slow). This does mean, however, that it is {\em not} possible to reconcile 
the observations with a model in which galaxy assembly uniformly tracks halo assembly, 
even allowing for the final galaxy stellar-halo mass ratio to be a function of halo mass
(i.e.\ setting in all progenitors the effective $M/L$ of the $z=0$ halo, which then 
simply assembles). 

%\clearpage
%\clearpage

\section{Summary \&\ Discussion}
\label{sec:discussion}

We compile a large number of observations of red/elliptical galaxy 
mass functions, and use these to determine the rate of ``buildup'' of the 
red sequence (RS) as a function of mass and redshift. Comparing these with 
observations of other populations allows us to test a number of 
different models for the possible associations between these populations 
and the ``transition'' of galaxies from blue, star-forming disks to red, 
``dead'' ellipticals. 

Independent of the nature of ``downsizing'' in the buildup of 
RS mass functions (discussed below), the rate of RS ``buildup'' is sufficiently well-determined to 
place meaningful constraints on a number of models. 
Dissipationless (gas poor, red, or ``dry'') mergers can 
account for the buildup of the RS at only the largest masses 
$\gtrsim10^{11}\,M_{\sun}$ at low redshift ($z\lesssim0.3$). At higher 
redshifts ($z\gtrsim0.5$), the dry merger rate would have to be at least an order of magnitude 
larger than observationally estimated \citep{vanDokkum05,Masjedi06,Bell06,Bell06b,Lotz06b} to account 
for observed RS buildup, even at the highest masses. This is perhaps 
unsurprising, as these and other observations find the 
{\em gas-rich} merger rate/fraction is an order of magnitude or more larger at all but the lowest 
redshifts. Furthermore, the total mass density on the RS is observed to increase by a 
factor $\sim2.5-3$ since $z\sim1$ \citep[e.g.,][]{Bundy05.full,Franceschini06,Pannella06,Borch06}, 
and dry mergers cannot, by definition, move/form ``new'' galaxies and mass on the RS. 

However, we find the {\em total} observed merger population (gas rich+gas poor) agrees very well 
with that expected if all RS galaxies are formed in mergers. Both the detailed mass 
distribution and fraction/rates of galaxy mergers are consistent with 
the rate of RS buildup at all masses and redshifts observed. This merger population 
is dominated by gas-rich mergers at all masses at high redshifts \citep[$z\gtrsim0.5$][]{Bell06b,Lotz06b} 
and at low masses at low redshifts, 
morphologically identifiable as bright (i.e.\ star-forming or starbursting) 
interacting systems \citep[e.g.,][]{Bundy05.masslimit,Wolf05}. In detail, 
completely neglecting dry mergers (or merger mass functions sensitive to them), 
this agreement is unchanged except for the highest masses at low-$z$ discussed above. 
There is substantial systematic uncertainty in converting a merger fraction 
to a merger rate; our comparisons assume a characteristic observable merger 
timescale $\sim0.5\,$Gyr. However, this is a theoretically reasonable timescale 
(see \S~\ref{sec:compare.merger.mass}), and given the scatter in the observations, our conclusions 
are not changed for systematic shifts within a factor $\lesssim2$, nor for 
allowing the merger timescale to scale with halo dynamical times ($\propto(1+z)^{-3/2}$). 
Furthermore, this has no effect our comparison of the mass distributions of these populations. 

Similarly, we find the rate at which host galaxies trigger quasars, determined 
as a function of the host stellar 
mass and redshift from the quasar luminosity function, agrees well with the observed RS buildup 
at all masses and redshifts observed. There is some discrepancy at the lowest redshifts 
and highest masses, but this is again where the dry merger contribution can account 
for the observed buildup, and dry mergers 
(by definition being gas-poor or gas-free) are not   
expected to trigger quasar activity. We consider this comparison first in a purely empirical 
fashion, using observed quasar Eddington ratios and the black hole-host mass 
relation to estimate quasar host masses as a function of redshift, and then in greater 
detail adopting the models of quasar light curves and lifetimes as a function of luminosity 
and host properties from the simulations of merger-induced quasar activity 
in \citet{H06b}. 
The latter introduces some model dependence (although it is consistent with the 
Eddington ratio and black hole-host mass relation estimates), but allows us to 
consider this comparison in greater detail and to make specific predictions for the 
characteristic host masses of quasars as a function of their position on the QLF and 
for the AGN or ``active'' fraction of galaxies as a function of stellar mass. In either case, 
the agreement between the rates of quasar formation/triggering as a function of 
host stellar mass and the buildup of RS galaxies is similar. 

We independently test these possible associations by comparing clustering 
measurements of the relevant populations as a function of redshift, and 
find similar results. The clustering of quasars and systems ``in transition'' to the RS agree at all 
redshifts as if they trace the same mass distribution. Clustering properties of 
merger (ULIRG and SMG) and post-merger (E+A) populations are consistent, but considerably 
less well-constrained. 

Although the above comparisons do not technically depend on it, we 
determine the ``transition'' mass ($\mtrans$, $M_{Q}$), i.e.\ the mass which separates the 
blue, star-forming disk and red, non star-forming elliptical populations, 
as a function of redshift. It has been suggested \citep[e.g.,][]{Bundy05.masslimit} that this 
represents the characteristic mass at which galaxies are forming on or being 
added to the RS as a function of redshift, but quantified in this manner, it is not 
obviously so \citep[see, e.g.][]{Shankar06}. We therefore also determine 
\citep[$M_{50}$, following][]{Cimatti06} 
the minimum mass above which the RS mass function 
is $\gtrsim50\%$ assembled at a given redshift. Regardless of definition, 
and furthermore regardless of the criterion used to separate early and late-type 
populations (whether e.g.\ a color, star formation rate, or morphology criterion), 
$\mtrans$/$M_{Q}$/$M_{50}$ shift to systematically larger masses 
at higher redshift (significant at $>6\sigma$), tracing a very similar trend 
as a function of redshift. 

This trend, especially in $M_{50}$ (which is independent of 
possible evolution in late-type mass functions) 
suggests that ``downsizing'' applies not just to galaxy {\em star formation}, 
but also in some sense to galaxy {\em assembly}, as suggested by the 
studies of e.g.\ \citet{Bundy05.full,Bundy05.masslimit,Zucca05,Yamada05,
Franceschini06,Cimatti06,Fontana06,Brown06}. 
In greater detail, considering the full rate of RS buildup as a function of stellar mass and 
redshift, low mass ($\lesssim10^{11}\,M_{\sun}$) 
galaxies appear to be building up rapidly/continuously at low redshifts ($\sim7-15\%$ per Gyr), 
but the most massive systems do not ($\sim1\%$ per Gyr growth at $z\sim0$). 
The growth of the most massive systems instead appears to be rapid at significantly 
higher redshifts (e.g.\ rising to $\sim20-50\%$ per Gyr by $z\sim1$). Equivalently, 
the characteristic mass (Schechter function $M_{\ast}$) defined by 
this ``formation'' rate appears (albeit at only $\sim2-3\,\sigma$) to increase with redshift in a similar fashion to the ``transition'' mass. 

We compare the ``transition'' mass with the characteristic masses of mergers and quasars
and again find they trace similar masses as a function of redshift, with ``downsizing'' evident in 
all three populations ($>3\sigma$ for mergers, $>6\sigma$ for quasars), further supported 
by their observed clustering. We compare 
with the characteristic (Schechter function) $M_{\ast}$ of the entire, red, and blue galaxy populations, 
and rule out at high significance the possibility that ``transition'' mass objects are drawn uniformly 
from any of these populations. Even with the systematic uncertainties in this mass estimate, 
it is also clearly distinct as a function of redshift from the characteristic masses of 
e.g.\ cluster, radio galaxy, ERO, DRG, or LBG populations \citep[see e.g.\ Figure~2 of][]{Farrah06}. 

These observations are all consistent with and suggest a scenario in which 
major mergers, quasars, and the transition from blue disk to red elliptical galaxies are associated. 
They do not inform us regarding, for example, whether gas exhaustion or stellar or quasar 
feedback is the specific mechanism for the reddening which accompanies the 
merger-driven morphological transformation and quasar episode. However, they 
support the hypothesis that mergers drive the transition from blue disks to
red elliptical galaxies, terminating in decaying, feedback-driven
bright quasar phases. The transition mass and break in the QLF appear to reflect
the characteristic mass of gas-rich objects merging at a given
redshift, which may build up the {\em new mass} on the red sequence at progressively lower masses
at lower $z$ as gas supplies are exhausted in more massive systems. 

That quasar host masses trace the ``transition'' mass and not, e.g.\ the blue galaxy population $M_{\ast}$ 
(see also Figure~\ref{fig:compare.models}) rules out the possibility that quasar activity {\em generically} 
traces star formation, as variants of e.g.\ the \citet{Granato04} models might predict. 
Likewise, that quasar masses do not trace the red galaxy $M_{\ast}$, as they would if 
quasar activity was long-lived or randomly (but uniformly as a function of mass) 
episodic in high-mass black holes. The former case would be expected from the low-level AGN 
activity invoked in e.g.\ \citet{Croton05}, if ``radio'' and optical or high-Eddington ratio quasar activity were 
associated, but they are in fact generally believed to be distinct \citep[e.g.,][]{Ho02,White06,Koerding06}. 
The latter case implies strong limits on implementations of e.g.\ the 
\citet{Binney04} model, which seek to suppress cooling flows through sporadic but potentially 
high accretion rate AGN activity. 

Although we can rule out some alternatives to the merger scenario, 
there remain a number of viable variants of ``quenching'' models, in which 
crossing the critical halo mass $\mhaloquench$ and entering a ``hot'' accretion regime plays a 
key role in the ``transition'' to the RS. Especially given 
that some feedback mechanism is typically required, even in the ``hot'' accretion regime, 
to prevent the formation of cooling flows, it is easy to imagine a scenario in which, upon 
entering this regime, new infalling halo gas is shock-heated, but feedback from e.g.\ a 
central disk galaxy with a small black hole is inefficient, cold gas reservoirs remain large, and 
cooling flows can form. Thus the system will not redden until it subsequently 
undergoes a major merger, which morphologically transforms the system, rapidly exhausts the 
remaining cold gas reservoir, and triggers a quasar and builds up a massive black hole, injecting 
some level of feedback and enabling efficient future (e.g.\ cyclic AGN or radio-mode) feedback. 
The ``hot'' accretion regime may be a necessary 
prerequisite for feedback to efficiently prevent subsequent cooling, and as discussed in 
\S~\ref{sec:halomass}, our comparisons are all consistent with this possibility. 
This is generally similar to the scenario assumed in e.g.\ \citet{Croton05}, although they 
do explicitly incorporate quasar light curves or feedback. Recognizing 
these distinctions (as opposed to e.g.\ assuming a system simply ``shuts down'' upon 
reaching $\mhaloquench$) will probably have little effect on the $z=0$ predictions 
of semi-analytic models, since in either case star formation will be effectively 
suppressed at relatively early times in the most massive systems \citep[see also][]{Cattaneo06}. 
However, at higher redshifts when massive objects 
are still forming, the distinctions will almost certainly be significant. 

We note that none of our conclusions conflict with the hypothesis that, once formed, 
elliptical galaxies can continue to grow by dry mergers. However, they emphasize that the 
importance of such mergers is restricted to the most massive galaxies at low redshifts. 
Our results, even the steep evolution of $M_{50}$ implying some ``downsizing'' in 
red galaxy assembly, are all consistent with (and in fact, marginally {\em favor}) the relatively low 
observationally inferred dry merger rate ($\sim1$ major dry merger since $z\sim1$). 
Essentially, ``downsizing'' in galaxy assembly as we have quantified it is 
not, strictly speaking, ``anti-hierarchical.'' Massive galaxies still continue to build up their 
populations to the present; it is simply a statement that the {\em relative} rates of 
red galaxy formation/assembly decrease or ``slow down'' at late times in the 
most massive systems. This could be related to pure dark matter processes, 
for example the rapid evolution in large overdensities could simply exhaust 
the ``supply'' of galaxies with which to merge, or the cluster environments of 
massive systems at low redshift attain sufficient circular velocities as to 
rapidly reduce merger rates \citep[see also][]{Neistein06}. 
The evolution of the ``transition'' mass may, alternatively, 
be a statement that galaxy assembly 
does not strictly trace halo assembly. There are a number of 
baryonic processes which make this possible, as it simply requires that 
the effective baryon conversion efficiencies in galaxies be a 
function of time, or different for central vs.\ satellite systems. 

Improved measurements of early-type mass functions at high redshift ($z\sim1$), 
larger samples of mergers from which to construct merger mass functions, 
revised or direct determinations of high-redshift conditional mass functions, 
and direct observations of the masses of quasar hosts 
will substantially improve the constraints in this paper. Ultimately, the integration of 
the merger and quasar host mass functions may enable a purely 
observational comparison with the remnant, red galaxy mass function. 
Calibration of the observable merger ``timescale'' with realistic
high resolution galaxy merger simulations, i.e.\ calibration of selection efficiencies 
for observed merger fractions, can further remove the factor $\sim2$ uncertainty 
in comparing the rates of elliptical buildup and observed merger populations. 
The association favored here between mergers, quasars, and elliptical 
buildup also makes specific predictions for the characteristic masses of 
E+A galaxies and quasar hosts as a function of redshift, which should be 
testable in future wide-field surveys.  

The scenario we have described does not, of course, imply 
that mergers, quasars, and remnant ellipticals will necessarily 
be recognizable as the same, singular objects at a given instant -- in fact, simulations 
which follow the transition through these stages \citep[e.g.,][]{H06c}, 
predict that they will be seen as distinct phases in merger-triggered 
evolution, and observations tracking e.g.\ the 
associations between dynamical merger state and quasar activity 
\citep[e.g.,][]{Straughn05} support this distinction. 
What we ultimately find evidence for here 
in the masses, luminosities, and clustering properties of 
mergers, galaxies being ``added'' or ``in transition'' to the red sequence, 
and quasars is that they are 
drawn from the same ``parent'' population, and that this population 
is distinct from the ``quiescent'' all/red/blue galaxy population. Again, none of 
this strictly implies causality, but it does favor models which 
associate these populations with the same event, a natural 
expectation if mergers of gas-rich galaxies trigger quasars and 
morphologically transform disks to spheroids, moving ``new'' 
mass to the early-type population and leaving an elliptical, 
gas-poor, rapidly reddening remnant galaxy.

\acknowledgments We thank Eric Bell, Arjun Dey, David Hogg, 
Casey Papovich, Rachel Somerville, Stijn Wuyts, Sandy 
Faber, and Marijn Franx for very helpful discussions. We also 
thank T.~J.\ Cox, Brant Robertson, and the anonymous referee whose 
comments improved this manuscript. 
This work was supported in part by NSF grants ACI
96-19019, AST 00-71019, AST 02-06299, and AST 03-07690, and NASA ATP
grants NAG5-12140, NAG5-13292, and NAG5-13381.

\end{document}